\documentclass{amia}
\usepackage{lipsum} 
\usepackage{rotating}
\usepackage{multirow}
\usepackage{color}
\usepackage{threeparttable}
\usepackage{graphicx}
\usepackage{subcaption}
\usepackage{amsmath}
\usepackage{xcolor}
\usepackage{hyperref}
\usepackage{setspace} 
\usepackage{comment}
\usepackage{tabularx}   
\usepackage{makecell}

\begin{document}
\pagenumbering{arabic}
\pagestyle{plain}

\newpage
\title{Enhancing End Stage Renal Disease Outcome Prediction: A Multi-Sourced Data-Driven Approach}

\author{Yubo Li, MS$^1$, Rema Padman, PhD$^1$}

\institutes{
    $^1$ Carnegie Mellon University, Pittsburgh, PA, USA
}

\maketitle

\section*{Abstract}
\vspace{-.3cm}

\textbf{Objective}: To improve prediction of Chronic Kidney Disease progression to End Stage Renal Disease using machine learning and deep learning models applied to integrated clinical and claims data with varying observation windows, supported by explainable AI to enhance interpretability and reduce bias.

\vspace{-.3cm}
\textbf{Materials and Methods}: We utilized data from 10,326 CKD patients, combining clinical and claims information from 2009-2018. After preprocessing, cohort identification, and feature engineering, we evaluated multiple statistical, ML and DL models using five distinct observation windows. Feature importance and SHAP analysis were employed to understand key predictors. Models were tested for robustness, clinical relevance, misclassification patterns, and bias.

\vspace{-.3cm}
\textbf{Results}: Integrated data models outperformed single data source models, with Long Short-Term Memory (LSTM) achieving the highest AUC (0.93) and F1 score (0.65). A 24-month observation window optimally balanced early detection and prediction accuracy. The 2021 eGFR equation improved prediction accuracy and reduced racial bias, particularly for African American patients.

\vspace{-.3cm}
\textbf{Discussion}: Improved prediction accuracy, interpretability and bias mitigation strategies have the potential to enhance CKD management, support targeted interventions, and reduce healthcare disparities.

\vspace{-.3cm}
\textbf{Conclusion}: This study presents a robust framework for predicting ESRD outcomes, improving clinical decision-making through integrated multi-sourced data and advanced analytics. Future research will expand data integration and extend this framework to other chronic diseases.

\vspace{-.3cm}

\section{Introduction}

\vspace{-.3cm}
Chronic Kidney Disease (CKD) is a complex, multi-morbid condition marked by a gradual decline in kidney function, which can ultimately progress to end-stage renal disease (ESRD) \cite{NKF_CKD_2024}. With a global prevalence of CKD ranging from $8\%$ to $16\%$, and estimates suggesting that around 5-10\% of individuals diagnosed with CKD eventually reach ESRD\cite{lin2013progression}, they represent a major public health challenge, particularly due to its strong associations with diabetes and hypertension \cite{NCHS2019Mortality}. CKD progression is classified into five stages, culminating in ESRD, where kidney function drops to $10\text{--}15\%$ of normal capacity, necessitating dialysis or transplantation for patient survival \cite{NKF_CKD_2024}. The economic impact of CKD is significant, with a relatively small proportion of Medicare CKD patients in the United States contributing to a disproportionately high share of Medicare expenses, particularly when they progress to ESRD. Additionally, more than one-third of ESRD patients are readmitted within 30 days of discharge, underscoring the critical need for early detection and management of CKD to prevent its progression to ESRD and to reduce healthcare costs \cite{guo2020machine}.


Previous CKD progression prediction efforts used either EHR clinical data \cite{belur2020machine} or administrative claims \cite{krishnamurthy2021machine, li2024towards}. These approaches often use limited features, not fully capturing CKD progression complexity. Sharma et al. \cite{sharma2020model} developed a claims-based model to identify CKD patients at Hyperkalemia risk, while Krishnamurthy et al. \cite{krishnamurthy2021machine} predicted CKD onset using comorbidities and medications. Claims data typically lacks clinical data's granularity. Tangri et al. \cite{tangri2011predictive} used age, gender, and eGFR \cite{NKF_eGFR} to predict ESRD progression, and Keane et al. \cite{sun2020development} utilized serum creatinine and urine protein levels for high-risk patient identification. Models using only clinical data may miss complete healthcare system interactions, struggle with missing data and inconsistent recording, and overlook socio-economic factors and healthcare utilization patterns, limiting their applicability across diverse populations.

This study bridges a critical gap by developing a framework that utilizes integrated clinical and claims data rather than isolated data sources. By minimizing the observation window needed for accurate predictions, our approach balances clinical relevance with patient-centered practicality. This integration enhances both predictive accuracy and clinical utility, enabling more informed decision-making to improve patient outcomes.

\vspace{-.3cm}

\section{Objective}

\vspace{-.3cm}

\begin{figure}[H]
\begin{center}
\includegraphics[width=1\textwidth]{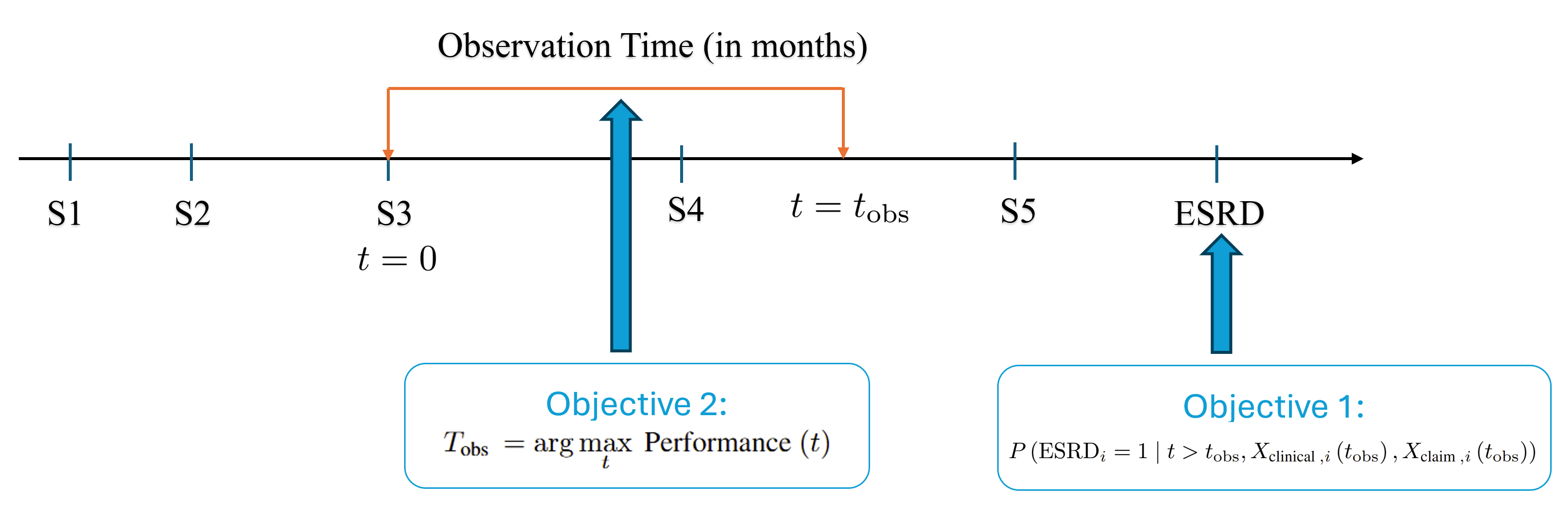}
\caption{A timeline illustration of observation windows for CKD progression to ESRD. The observation starts at CKD Stage 3. Objective 1 estimates the probability of ESRD occurrence after the observation window, using clinical and claims data within the observation window, while Objective 2 identifies the optimal observation window that maximizes predictive performance.}
\label{fig:timeline_stage}
\end{center}
\end{figure}

This research evaluates predictive models for CKD progression to ESRD using integrated clinical and claims data. Fig. \ref{fig:timeline_stage} illustrates our study design, where the observation window begins at the initial diagnosis of CKD Stage 3 (t=0). Although patients may progress through subsequent CKD stages (Stages 4 and 5) during this observation period, our cohort specifically excludes those developing ESRD within this timeframe. This ensures that predictive modeling utilizes only pre-ESRD data for forecasting future ESRD onset.
\begin{itemize}
    \item Our primary objective, ESRD risk prediction, aims to estimate the probability that a patient diagnosed with CKD Stage 3 will develop ESRD after the observation window, represented as follows:
    $$
    P\left(\mathrm{ESRD}_i=1 \mid t > t_{\mathrm{obs}}, X_{\text {clinical }, i}\left(t_{\mathrm{obs}}\right), X_{\text {claim }, i}\left(t_{\mathrm{obs}}\right)\right),
    $$
    
    where \( X_{\text{clinical}, i}\left(t_{\text{obs}}\right), X_{\text{claim}, i}\left(t_{\text{obs}}\right) \) denote the clinical and claims data observed up to time \( t_{\text{obs}} \) for patient \( i \).

    \item  To identify the optimal observation window $T_{\mathrm{obs}}$ that provides the best performance across all candidate windows $t_{\mathrm{obs}}$ (6, 12, 18, 24, and 30 months), ensuring accurate predictions while minimizing the length of the observation window required:

    $$
    T_{\text {obs }}=\arg \max _{t \in\{6,12,18,24,30\}} \text { Performance }(t),
    $$
 
    where Performance $(t)$ is evaluated based on key predictive metrics: F1-score \cite{vanRijsbergen1979}, Area Under the Receiver Operating Characteristic curve (AUROC) \cite{hanley1982meaning}, and Area Under the Precision-Recall Curve (AUPRC) \cite{davis2006pr}.
\end{itemize}

\section{Data Sources}

This study utilized two integrated datasets: administrative claims data and clinical data. The claims dataset includes patient healthcare interactions, spanning from 2009 to 2018, containing diagnosis codes, treatment records, and healthcare costs for individuals diagnosed with CKD. The clinical dataset from electronic health records contains laboratory results, patient demographics, diagnostic details, and medication records, which was truncated to match the 10-year span of the claims data for consistency of integration. See Appendix \ref{app:data_details} for data details.

\section{Methods}

\begin{figure}[htbp]
  \centering
  \begin{subfigure}[b]{\textwidth}
    \centering
    \includegraphics[width=0.75\textwidth]{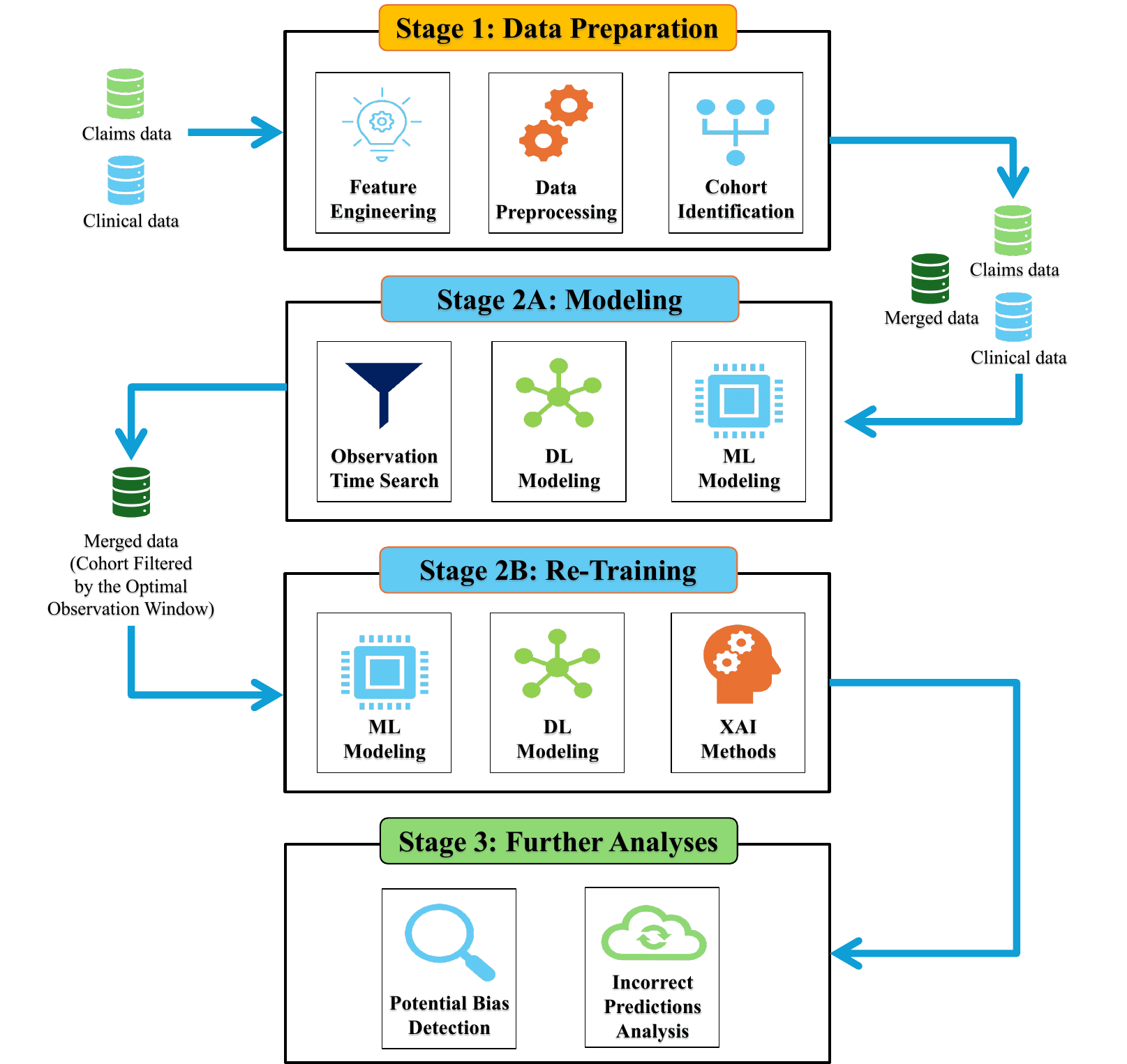}
    \caption{Three-stage pipeline for predicting CKD progression to ESRD. Stage 1: feature engineering, data preprocessing, and cohort identification using clinical and claims data. Stage 2: (2A) systematic evaluation of observation windows with ML/DL models; (2B) retraining the optimal model on the filtered cohort from the optimal observation window, followed by explainable AI analysis. Stage 3: Additional analyses including bias detection and misclassification assessment.}
    \label{fig:exp_pipeline}
  \end{subfigure}

  \vspace{1em}  

  \begin{subfigure}[b]{\textwidth}
    \centering
    \includegraphics[width=0.8\textwidth]{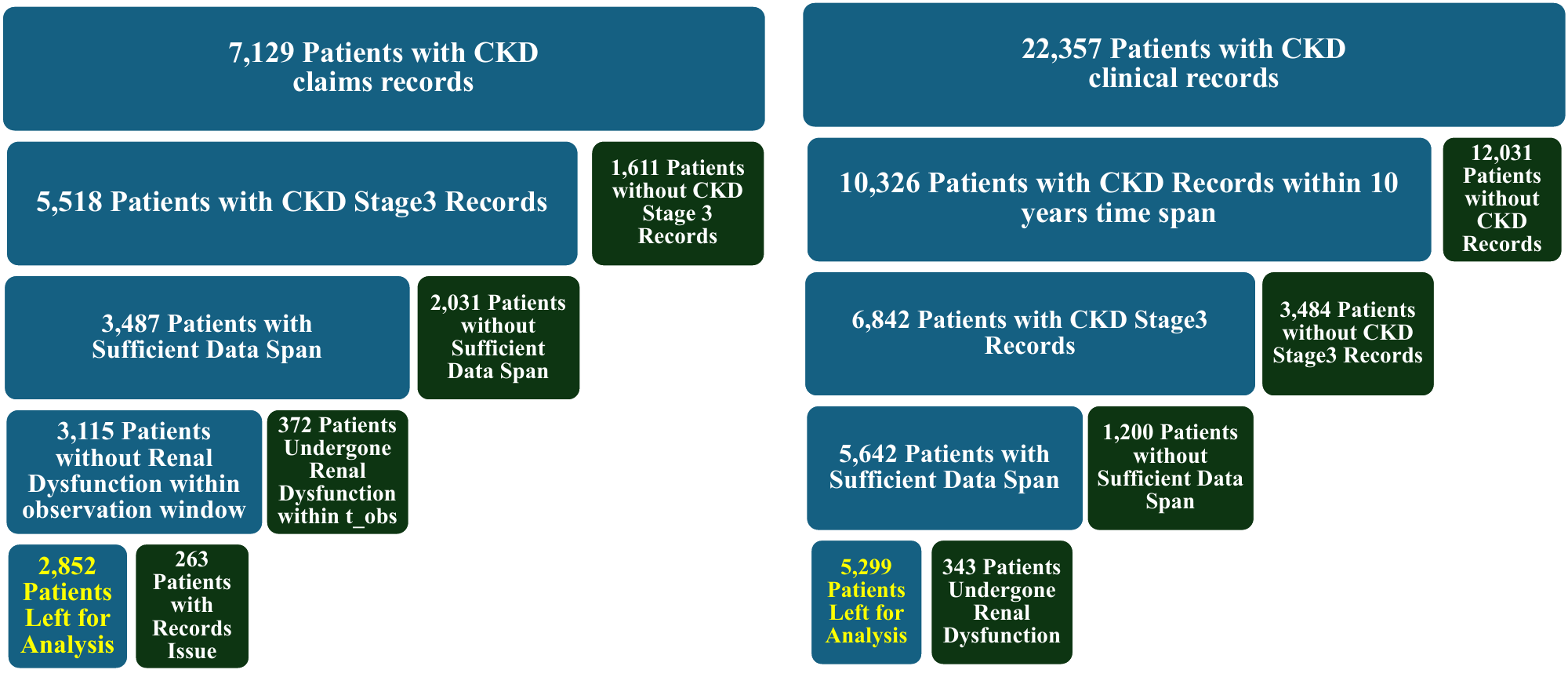}
    \caption{Multi-step cohort curation process for longitudinal CKD analysis. Starting with initial selection of patients at CKD stage 3, an adjustable observation window (18 months shown) was applied. Only patients with complete clinical and claims data throughout this period were retained, while those progressing to ESRD within the observation window were excluded, ensuring a cohort for analysis (highlighted in yellow) with uninterrupted longitudinal records.}
    \label{fig:cohort_identification}
  \end{subfigure}

  \caption{Framework for CKD progression prediction: (A) Overview of the three-stage analytical pipeline, and (B) Cohort identification methodology.}
  \label{fig:frameworks}
\end{figure}

As illustrated in Figure \ref{fig:exp_pipeline}, our comprehensive methodology for predicting CKD progression to ESRD consists of three primary stages: Data Preparation, Modeling, and Additional Analyses. Within the Modeling stage, we employ a two-phase analytical framework that first systematically evaluates various predictive models across multiple observation windows (but the same patient cohort for consistent comparisons)  to determine optimal performance, and subsequently retrains the most effective model on the cohort for the each observation window for practical clinical application. This framework bridges statistical rigor and clinical relevance, facilitating actionable insights for nephrologists managing patients with CKD stage 3 and above.

\subsection{Data Preparation}

\subsubsection{Data Pre-processing}

We removed duplicate records to prevent redundancy and bias in the analysis. Entries lacking CKD diagnoses were excluded to maintain relevance to study objectives, along with claims containing negative values (likely data entry errors). The cleaned claims dataset comprised 5,317,178 claims across 7,129 unique patients, while the refined clinical dataset included 433,421 laboratory records for 10,326 patients. These datasets were integrated using unique patient IDs to provide a comprehensive view of each patient's medical and claims history.

To address missing data, we employed multiple imputation via chained equations (MICE) \cite{white2011multiple}. Continuous numerical features were standardized, and skewed variables were log-transformed \cite{sedgwick2012log} to stabilize variance and mitigate outliers. For laboratory values missing unit specifications, we used distribution detection and adjustment \cite{aryal2018anomaly} to identify outliers and harmonize units across records (medical expert-advised ranges in Appendix \ref{app:ranges}).

\subsubsection{Cohort Identification}


As depicted in Fig.~\ref{fig:cohort_identification}, cohorts were curated through a multi‐step process: patients at CKD stage 3 were selected and an observation window—18 months in the example, but adjustable per study—was applied. Only individuals with complete clinical and claims data throughout this window were retained, and those who progressed to ESRD within it were excluded, yielding a cohort for analysis.

\subsubsection{Feature Engineering}

Feature engineering involved identifying predictive variables essential for modeling. For claims data, we derived two main groups of features: cost-based features, including claim counts, aggregate patient costs, cost ranges, and cost standard deviations from inpatient, outpatient, professional, pharmacy, and vision claims; and comorbidity-based features, such as CKD stage 3 duration, emergency department visit frequency, and critical comorbidities (e.g., hypertension, diabetes, phosphatemia).

Clinical data features were categorized based on nephrology expert consultations, relevant literature, and dataset characteristics into demographic features, including age, gender, ethnicity, and BMI as established CKD progression indicators; laboratory features comprising essential clinical parameters like eGFR, hemoglobin, phosphorus, serum calcium, and bicarbonate (excluding UACR~\cite{NIDDK_UACR} and sodium due to insufficient data availability); and additional comorbidity features, capturing conditions such as cardiovascular disease, anemia, and metabolic acidosis, which are known to influence CKD outcomes.

Statistical analyses compared variables between patients progressing to ESRD and those who did not. For continuous variables, data were assessed for normality; normally distributed variables were analyzed using independent t-tests to compare group means. Non-normally distributed variables underwent logarithmic transformation during preprocessing before applying t-test. Categorical variables were expressed as frequencies (percentages) and compared using chi-squared tests. Statistical significance was set at $p < 0.05$ for all analyses.

\subsection{Modeling \& Validation}


We trained both ML and DL models on stratified train/validation/test splits to preserve class proportions, and applied SMOTE \cite{chawla2002smote, saif2024early} to the training data to oversample the minority class.

\subsubsection{Machine Learning Methods}

We employed logistic regression (LR) \cite{kleinbaum2010logistic} as our baseline model for ESRD progression probability estimation. We extended our analysis to Random Forest (RF) \cite{breiman2001random} and XGBoost \cite{chen2016xgboost}, evaluating performance via k-fold cross-validation \cite{wong2019reliable}. RF utilizes multiple decision trees to enhance accuracy while reducing overfitting. XGBoost sequentially builds models to correct previous errors, proving highly effective for structured data.

\subsubsection{Deep Learning Methods}

To develop a robust predictive model for CKD progression to ESRD, we adopted methodologies from prior studies \cite{li2024towards, burckhardt} by constructing a longitudinal data representation with three-month intervals following initial CKD stage 3 diagnosis. These intervals were sequentially indexed as timestamps (0 for months 0-3, 1 for months 3-6, etc.).

This temporal segmentation approach serves several purposes, capturing CKD’s temporal evolution, reflecting its gradual progression, aligning with clinical guidelines recommending periodic nephrologist visits, and accommodating potential time lags between clinical events and their corresponding claims data. These lags, particularly notable for critical events such as transplantation or dialysis initiation, exhibit substantial variability both across and within patient records, complicating consistent correction strategies. Employing three-month intervals mitigates the impact of these temporal discrepancies, improving model accuracy. Numerical features were aggregated, categorical features encoded binarily to reflect conditions or events, and baseline historical records were analyzed to adjust for pre-existing conditions, thereby ensuring accurate representation of CKD progression. This timestamp-based structure allows models to capture disease progression dynamics and identify critical temporal patterns signaling heightened ESRD risk.


We evaluated multiple deep learning architectures, including Convolutional Neural Networks (CNN)\cite{lecun2015deep} for feature extraction; Recurrent Neural Networks (RNN)\cite{rumelhart1986learning}, Long Short-Term Memory networks (LSTM)\cite{hochreiter1997long}, and Gated Recurrent Units (GRU)\cite{cho2014learning} for modeling temporal dependencies; and Temporal Convolutional Networks (TCN)\cite{bai2018empirical} for sequence modeling. These architectures were selected to optimize prediction accuracy of ESRD progression and improve identification of associated risk factors. Detailed implementation information is available in Appendix \ref{app:implementation_details}.

To maximize clinical relevance, our analytical framework employs a two-phase approach (as illustrated in Fig.~\ref{fig:exp_pipeline} Stage 2). First, we systematically compare all modeling techniques mentioned above across multiple observation windows to identify optimal predictive performance. Second, recognizing that clinicians cannot predetermine a patient's progression timeline to ESRD in real-world settings, we re-train the best-performing model using the complete cohort with the optimal observation window. This clinically-oriented model then undergoes comprehensive explainable AI analysis to identify key predictive features and temporal patterns. This approach bridges the gap between statistical performance and practical clinical implementation, providing actionable insights for nephrologists. 

\subsection{Explainable AI Methods}
After identifying the best-performing model and optimal observation window, we employed two complementary explainable AI techniques to understand the features driving these predictions. At the cohort level, we utilized feature importance analysis to identify key predictive variables across the population. At the individual level, we applied SHAP analysis to provide detailed, patient-specific insights into model predictions. 

\subsection{Additional Analyses}

Beyond the core dual-phase predictive framework outlined above, we conducted several additional analyses to enhance the robustness and fairness of our ESRD prediction models. Specifically, we investigated cases of model misclassification to identify underlying patterns and contributing factors, providing deeper insights into prediction limitations. Furthermore, we assessed the influence of recently updated eGFR equation on racial disparities within CKD progression predictions, aiming to ensure clinical fairness and improve the accuracy of our predictive models.

\subsubsection{Analysis of Model Misclassifications}
We examined predictive probability distributions across our sample set to analyze misclassifications from our optimal model. We evaluated both false positives (type I errors) and false negatives (type II errors) in ESRD progression predictions to identify patterns contributing to incorrect predictions. This error analysis provides insights into model limitations and highlights potential areas for refinement, ultimately improving clinical reliability and decision support.

\subsubsection{Impact of Updated eGFR Equations on Racial Bias in CKD Predictions}
The 2021 workgroup led by the National Kidney Foundation and American Society of Nephrology recommended an updated CKD-EPI equation that removed race coefficients, addressing concerns about racial bias in eGFR calculations \cite{inker2021new, miller2022national}, such as race-based components lacking biological basis and risking the perpetuation of healthcare disparities \cite{levey2020kidney, diao2021clinical}.

For our study using data from 2009-2018, we primarily used the 2009 CKD-EPI equation since this was the standard clinical calculation during the study period and informed actual physician decisions and CKD staging. We additionally applied the 2021 race-free equation in our further analysis section to compare outcomes and assess improvements in prediction accuracy and equity, particularly for minority populations. Note that adopting the updated equation may affect CKD stage classification, potentially altering cohort composition and model predictions.

\section{Results}

\subsection{Cohort for Analysis \& Data Characteristics}

\begin{figure}[htbp]
  \centering
  \begin{subfigure}[b]{0.49\textwidth}
    \includegraphics[width=\textwidth]{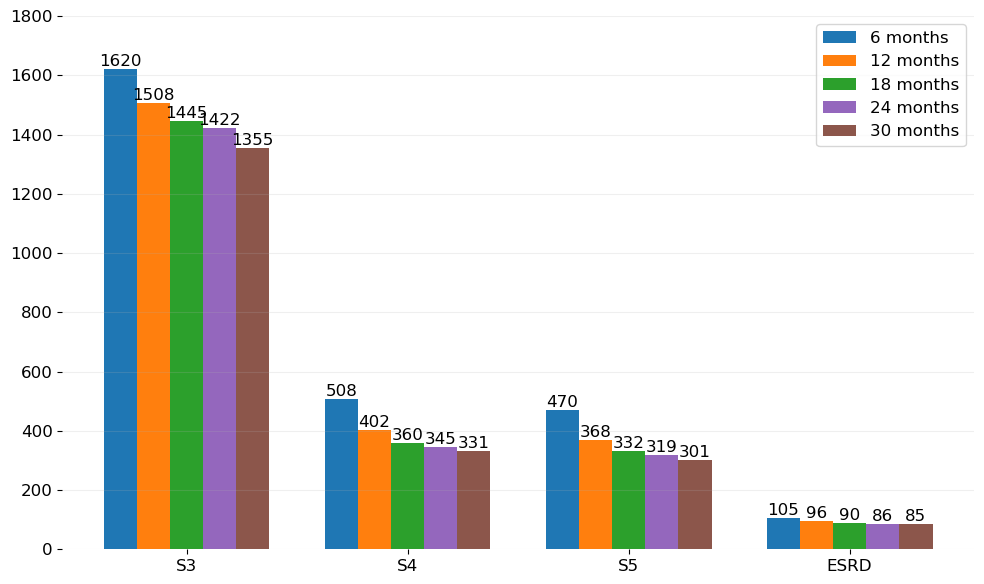}
    \subcaption{Variation in CKD stage distribution of merged data as a function of different observation window lengths.}
    \label{fig:merged_data_info_1}
  \end{subfigure}
  \hfill
  \begin{subfigure}[b]{0.49\textwidth}
    \includegraphics[width=\textwidth]{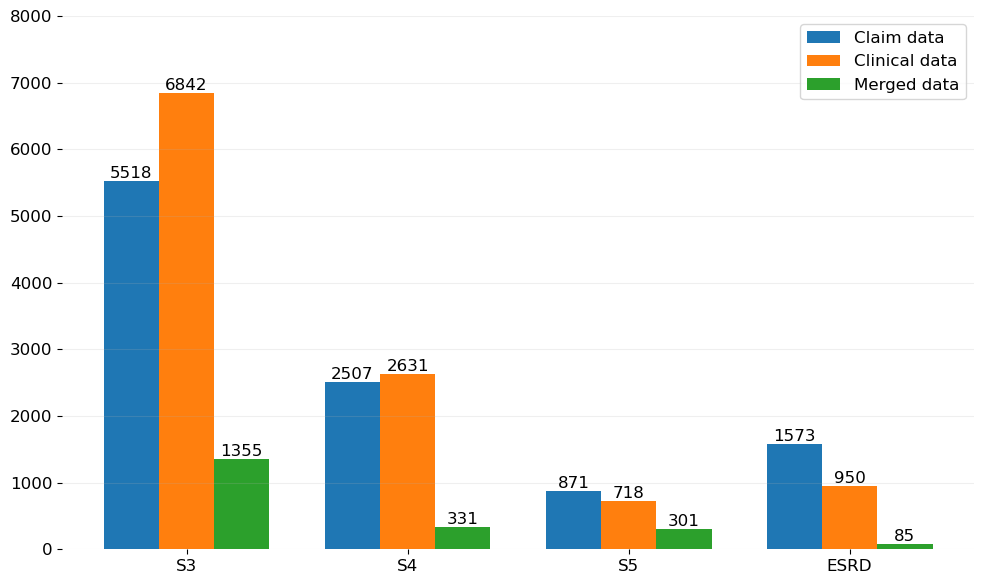}
    \subcaption{Comparison of CKD stage distribution across claims, clinical, and merged data with a 30-month observation window.}
    \label{fig:merged_data_info_2}
  \end{subfigure}
  \vspace{.2cm}
  \caption{Overview of CKD Stage Distribution in Merged Data Across Observation Windows.}
  \label{fig:merged_data_info}
\end{figure}

In this subsection, we present data integration trends and cohort characteristics. Fig.\ref{fig:merged_data_info_1} shows patient distribution across CKD stages with varying observation windows. Patient numbers decline as windows extend from 6 to 30 months due to insufficient data or ESRD progression within the window. The ESRD columns represent patients who progressed to ESRD after their respective observation windows. While some patient losses may be attributed to mortality, we lack confirmatory data.

To ensure model consistency and fair comparability, we trained and tested our models using the same patient cohort. Recognizing the issue of cohort sizes decreasing as the observation window lengthens, we selected patients who had at least 30 months of follow-up data after their CKD stage 3 diagnosis. We specifically chose 30 months because it represents the smallest cohort across the range of observation windows; thus, patients in this 30-month cohort are also guaranteed to be present in all shorter observation windows. For the training datasets, we utilized this same group of patients but varied the length of the data included according to different observation windows. This approach allowed us to systematically assess how different observation windows affected model performance and to identify the optimal window based on test dataset results.

After requiring $\ge$ 30 months of post–stage 3 data, we compared claim-only (n = 5,518) and clinical-only (n = 6,842) cohorts to the merged sample. Combining data reduced the cohort to 1,422 patients for the 24-month window (optimal) and 1,355 for the 30-month window (Fig.~\ref{fig:merged_data_info_2}), reflecting exclusion of those lacking one data source and yielding a more comprehensive dataset.

\begin{table}[htbp]
\setlength{\arrayrulewidth}{1pt}  
\caption{Data characteristics of the patient cohort for analysis (Merged data, 24-month observation window, n=1,422). Continuous variables are shown as mean ± standard deviation and compared by independent t-test (log-transformed if non-normal); categorical variables are shown as count (percentage) and compared by chi-squared test. “Missing data (\%)” indicates the proportion of missing observations for each variable. P-values are two-sided, with $p<0.05$ denoting statistical significance.}
\label{table:data_overview}
\small
\begin{tabular}{lcccc}
\hline
Characteristics & Missing data (\%) & Progressed to ESRD (n = 86) & Non-progressed to ESRD & P-value \\
& & & (n = 1,336) & \\
\hline
\textbf{Demographic} & & & & \\
Age (years) & 0 & 69.13 ± 12.37 & 72.04 ± 11.25 & $<$0.001 \\
Female & 0 & 40 (46.5\%) & 721 (54.0\%) & 0.2149 \\
Race & 0 & & & $<$0.001 \\
\hspace{0.2cm}White & & 70 (81.4\%) & 1242 (93.0\%) & \\
\hspace{0.2cm}African American & & 12 (14.0\%)  & 60 (4.5\%) & \\
\hspace{0.2cm}Others & & 4 (4.6\%) & 34 (2.5\%) & \\
BMI & 4 & 28.40 ± 5.32 & 26.40 ± 6.20 & $<$0.001 \\

\textbf{Comorbidities} & & & & \\
Hypertension & 0 & 85 (99\%) & 1,323 (99\%) & 0.863 \\
Diabetes & 0 & 63 (73.3\%) & 788 (59.0\%) & 0.009 \\
Anemia & 0 & 55 (64.0\%) & 828 (62.0\%) & 0.714 \\
Metabolic acidosis & 0 & 22 (25.6\%) & 240 (18.0\%) & 0.077 \\
Proteinuria & 0 & 11 (12.8\%) & 227 (17.0\%) & 0.312 \\
Secondary hyperparathyroidism & 0 & 28 (32.6\%) & 240 (18.0\%) & $<$0.001 \\
Phosphatemia & 0 & 4 (4.7\%) & 40 (3.0\%) & 0.39 \\
Heart failure & 0 & 6 (7.0\%) & 120 (9.0\%) & 0.526 \\
Stroke & 0 & 1 (1.2\%) & 40 (3.0\%) & 0.506 \\
Conduction \& dysrhythmias & 0 & 4 (4.7\%) & 214 (16.0\%) & 0.005 \\

\textbf{Claims-driven features} & & & & \\
Count of pharmacy claims & 0 & 120 ± 94 & 109 ± 86 & 0.293 \\
Count of inpatient claims & 0 & 3.85 ± 3.41 & 3.74 ± 3.62 & 0.773 \\
Count of outpatient claims & 0 & 27.78 ± 24.75 & 22.07 ± 19.13 & 0.039 \\
Count of professional claims & 0 & 105.37 ± 77.56 & 87.43 ± 68.02 & 0.039 \\
Net cost of pharmacy claims & 0 & 12053 ± 11596 & 10440 ± 20662 & 0.242 \\
Net cost of inpatient claims & 0 & 33909 ± 53540 & 29440 ± 32541 & 0.446 \\
Net cost of outpatient claims & 0 & 9354 ± 17522 & 8554 ± 17492 & 0.682 \\
Net cost of professional claims & 0 & 15512 ± 18657 & 11640 ± 12748 & 0.061 \\
Range of claims costs & 0 & 11352 ± 32606 & 8852 ± 11550 & 0.481 \\
Standard deviation of claims costs & 0 & 831 ± 1263 & 757 ± 806 & 0.593 \\

\textbf{Clinical-driven features} & & & & \\
eGFR & 0 & 17.21 ± 5.46 & 22.78 ± 5.66 & $<$0.001 \\
Hemoglobin & 3 & 12.15 ± 2.19 & 14.25 ± 1.8 & $<$0.001 \\

Bicarbonate & 9 & 22.9 ± 6.36 & 25.3 ± 4.22 & 0.001 \\
Serum calcium & 6 & 9.39 ± 3.62 & 10.21 ± 2.86 & 0.042 \\
Phosphorus & 13 & 3.61 ± 0.87 & 3.52 ± 0.72 & 0.350 \\

CKD stage 3 duration & 0 & 3.7 ± 0.6 & 3.9 ± 1.4 & 0.0009 \\
Occurrence of CKD stage 4 & 0 & 47 (54.7\%) & 298 (22.3\%) &  $<$0.001 \\
Occurrence of CKD stage 5 & 0 & 42 (48.8\%) & 277 (20.7\%) & $<$0.001 \\
Number of emergency department visits & 16 & 2.63 ± 2.18 & 2.01 ± 1.99 & 0.005 \\
\hline
\end{tabular}
\end{table}



Table~\ref{table:data_overview} summarizes characteristics of our 1,422-patient cohort; 86 (6\%) progressed to ESRD while 1,336 did not. Mean progression time from CKD stage 3 to ESRD was $4.82 \pm 1.82$ years. ESRD patients were younger ($69.13$ vs. $72.04$ years, $p<0.001$) with significant racial disparities: African Americans showed higher progression rates (14.0\% vs. 4.5\%) while White patients had lower rates (81.4\% vs. 93.0\%, $p<0.001$).

Hypertension was prevalent in both groups (99\%), but diabetes was more common in ESRD patients (73.3\% vs. 59.0\%, $p=0.009$). Secondary hyperparathyroidism and conduction disorders were also significantly higher in the ESRD group. Claims data showed slightly higher outpatient/professional claims counts for ESRD patients ($p=0.039$) but no significant cost differences. Clinically, ESRD patients had lower eGFR ($17.21$ vs. $22.78$, $p<0.001$), lower hemoglobin ($12.15$ vs. $14.25$, $p<0.001$), and more advanced CKD stages.

\subsection{Performance Comparison of Models Across Datasets}

To facilitate comprehension of model performance across different data sources, we present results using a 24-month observation window, which yielded optimal performance across most models. Table~\ref{table:performance_combined} compares the predictive performance of traditional machine learning and deep learning models across three scenarios: claims data-only, clinical data-only, and merged data.

The results demonstrate several key findings across the different data sources. In the claims data-only scenario, deep learning models, particularly LSTM (AUROC: 0.92, F1: 0.54) and GRU (AUROC: 0.92, F1: 0.50), substantially outperformed traditional machine learning approaches such as logistic regression (AUROC: 0.72, F1: 0.33) and random forest (AUROC: 0.74, F1: 0.36). For clinical data, while the performance gap narrowed, deep learning models maintained their advantage, with LSTM achieving the highest performance (AUROC: 0.88, F1: 0.60). Most notably, the merged dataset combining both claims and clinical data yielded the best overall performance, with LSTM achieving better results across all metrics (AUROC: 0.93, F1: 0.65, AUPRC: 0.61). Comprehensive results across observation windows are provided in Appendix~\ref{app:performance}.

\vspace{.3cm}
\begin{table}[htbp]
\caption{Model performance for prediction of ESRD, using claims data-only, clinical data-only, and merged data (24-month observation window, n=1,422).  \textbf{Bold} indicates the best performance within each method category (Machine Learning or Deep Learning).}
\label{table:performance_combined}
\centering
\begin{subtable}{0.8\textwidth}
  \centering
  \caption{Model performance for prediction of ESRD by using claims data.}
  \label{table:performance_1}
  \begin{tabular}{llccc}
    \hline
    \textbf{Model Type} & \textbf{Model} & \multicolumn{3}{c}{\textbf{Model performance metric}} \\
    \cline{3-5}
    & & F1 Score & AUROC & AUPRC \\
    \hline
    \multirow{3}{*}{\makecell{Machine Learning}}
      & Logistic Regression & 0.33 & 0.72 & 0.44 \\
      & Random Forest        & 0.36 & 0.74 & \textbf{0.48} \\
      & XGBoost              & \textbf{0.39} & \textbf{0.75} & 0.47 \\
    \hline
    \multirow{5}{*}{\makecell{Deep Learning}}
      & CNN                  & 0.45 & 0.82 & 0.50 \\
      & RNN                  & 0.50 & 0.90 & 0.52 \\
      & LSTM                 & \textbf{0.54} & \textbf{0.92} & \textbf{0.55} \\
      & GRU                  & 0.50 & \textbf{0.92} & 0.53 \\
      & TCN                  & 0.52 & 0.88 & 0.53 \\
    \hline
  \end{tabular}
\end{subtable}

\vspace{1cm}

\begin{subtable}{0.8\textwidth}
  \centering
  \caption{Model performance for prediction of ESRD by using clinical data.}
  \label{table:performance_2}
  \begin{tabular}{llccc}
    \hline
    \textbf{Model Type} & \textbf{Model} & \multicolumn{3}{c}{\textbf{Model performance metric}} \\
    \cline{3-5}
    & & F1 Score & AUROC & AUPRC \\
    \hline
    \multirow{3}{*}{\makecell{Machine Learning}}
      & Logistic Regression & 0.54 & 0.76 & 0.47 \\
      & Random Forest        & \textbf{0.58} & 0.79 & 0.51 \\
      & XGBoost              & 0.57 & \textbf{0.80} & \textbf{0.52} \\
    \hline
    \multirow{5}{*}{\makecell{Deep Learning}}
      & CNN                  & 0.56 & 0.84 & 0.53 \\
      & RNN                  & \textbf{0.61} & 0.85 & 0.53 \\
      & LSTM                 & 0.60 & \textbf{0.88} & \textbf{0.56} \\
      & GRU                  & 0.60 & 0.87 & 0.55 \\
      & TCN                  & \textbf{0.61} & 0.83 & 0.54 \\
    \hline
  \end{tabular}
\end{subtable}

\vspace{1cm}

\begin{subtable}{0.8\textwidth}
  \centering
  \caption{Model performance for prediction of ESRD by using merged data.}
  \label{table:performance_3}
  \begin{tabular}{llccc}
    \hline
    \textbf{Model Type} & \textbf{Model} & \multicolumn{3}{c}{\textbf{Model performance metric}} \\
    \cline{3-5}
    & & F1 Score & AUROC & AUPRC \\
    \hline
    \multirow{3}{*}{\makecell{Machine Learning}}
      & Logistic Regression & 0.55 & 0.75 & 0.45 \\
      & Random Forest        & 0.60 & 0.84 & 0.49 \\
      & XGBoost              & \textbf{0.61} & \textbf{0.85} & \textbf{0.51} \\
    \hline
    \multirow{5}{*}{\makecell{Deep Learning}}
      & CNN                  & 0.56 & 0.80 & 0.46 \\
      & RNN                  & 0.62 & 0.87 & 0.53 \\
      & LSTM                 & \textbf{0.65} & \textbf{0.93} & \textbf{0.61} \\
      & GRU                  & 0.63 & 0.90 & 0.58 \\
      & TCN                  & 0.61 & 0.89 & 0.58 \\
    \hline
  \end{tabular}
\end{subtable}
\end{table}

\subsection{Key Feature Analysis Using Explainable AI Techniques}

Leveraging the optimal 24-month window and our best ESRD prediction model, we performed (1) cohort-level feature importance to pinpoint the most influential predictors across the population, and (2) individual-level SHAP analysis to generate patient-specific explanations of each prediction.

\subsubsection{Cohort-Level Key Feature Identification}

\begin{figure}[htbp]
  \centering
  \begin{subfigure}[b]{0.44\textwidth}
    \includegraphics[width=\textwidth]{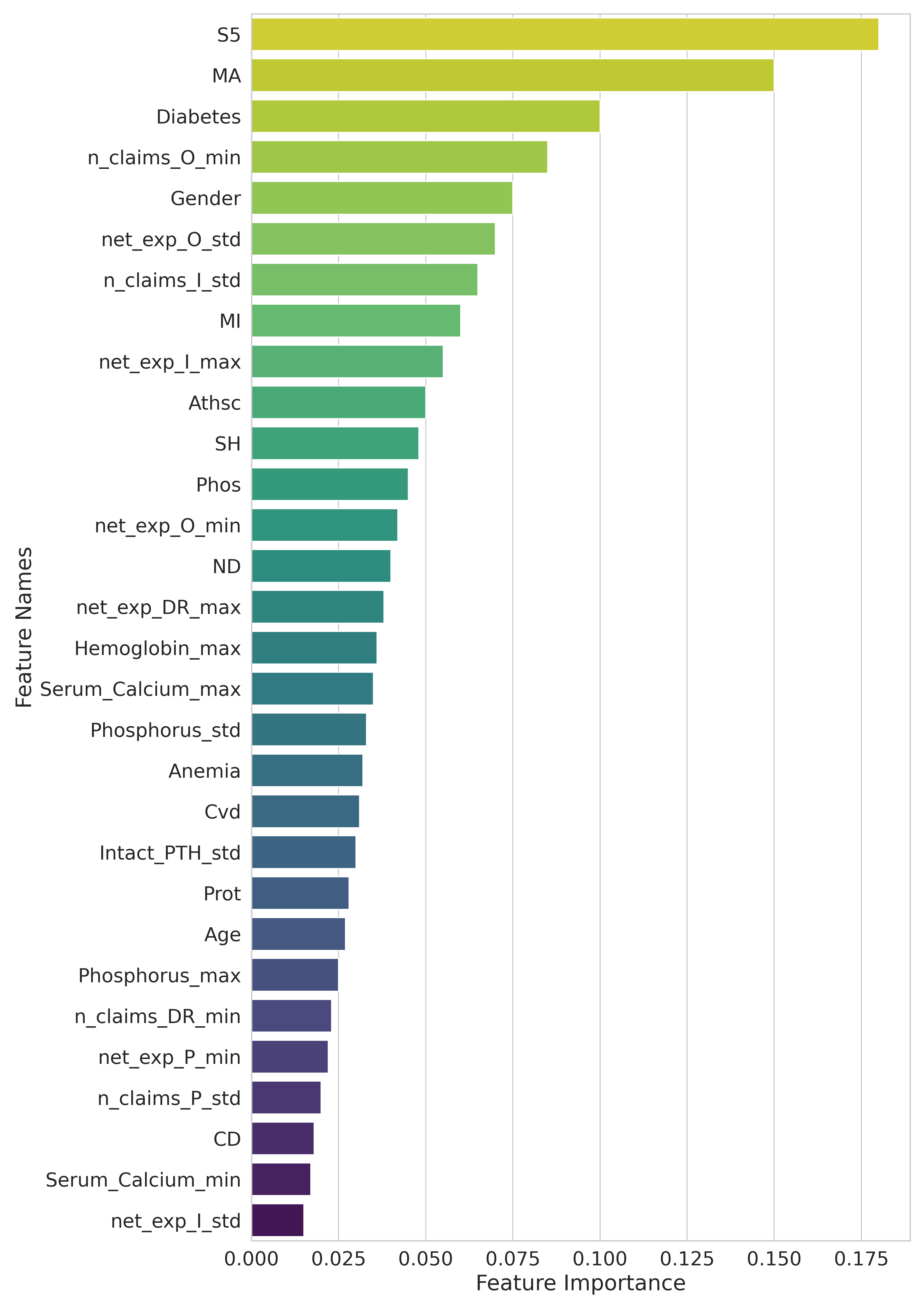}
    \subcaption{Top 30 features importance for the XGBoost model. Features are colored from yellow-green (highest importance) to dark blue (lower importance).}
    \label{fig:fea_imp}
  \end{subfigure}
  \hfill
  \begin{subfigure}[b]{0.55\textwidth}
    \includegraphics[width=\textwidth]{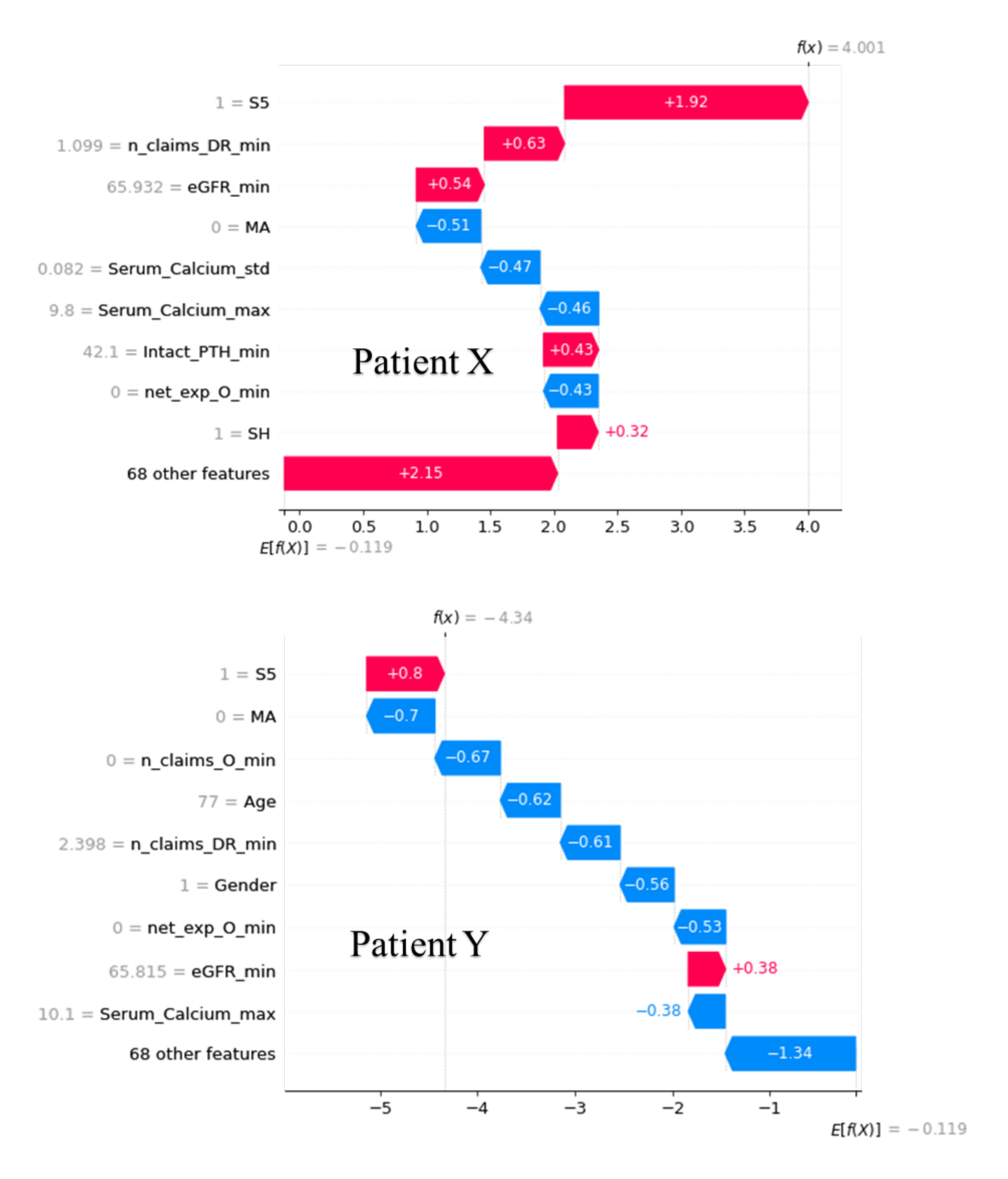}
    \subcaption{SHAP analysis demonstrating the variation in feature impact on ESRD risk prediction for two patients with CKD Stage 5. Red bars indicate features that increase predicted risk, while blue bars indicate features that decrease predicted risk.}
    \label{fig:shap}
  \end{subfigure}
  \vspace{.2cm}
  \caption{Feature importance and SHAP analysis for ESRD risk prediction using XGBoost model (24-month observation window, n=1,422). See Appendix~\ref{app:features_abb} for complete feature names.}
  \label{fig:key_feature_info}
\end{figure}

Our feature importance analysis of the optimal ML model, XGBoost, revealed several key predictors for ESRD progression (see Fig.~\ref{fig:fea_imp}). The most critical feature was the presence of CKD Stage 5  (S5), aligning with clinical literature and expert feedback on its importance in predicting ESRD. Claims-related features, such as the number of outpatient claims (n\_claims\_O) and total inpatient claim expenses (net\_exp\_I), also ranked highly. The diverse types of top-ranked features underscore the benefit of multi-sourced data integration, enhancing the model's predictive power. This combination of clinical and claims data not only improves the accuracy of the predictions but also supports more comprehensive and reliable decision-making in clinical practice. However, while XGBoost's feature importance reflects the absolute contribution of each feature, it does not specify whether the influence is positive or negative.

\subsubsection{Feature Impact at the Individual Patient Level Using SHAP Analysis}

To support personalized decision-making, we applied SHAP analysis to quantify the features driving each patient’s risk prediction. Figure~\ref{fig:shap} presents SHAP force plots for two correctly predicted high-risk patients, Patient X and Patient Y, both diagnosed with CKD Stage 5 but exhibiting distinct risk profiles.

For Patient X, elevated risk is driven primarily by Stage 5 CKD (S5) and a high volume of outpatient claims (n\_claims\_DR\_min), while higher eGFR levels and the absence of certain clinical markers mitigate that risk. In contrast, Patient Y’s lower risk contribution—despite the same Stage 5 diagnosis—stems from younger age, fewer minimum inpatient claims (net\_exp\_O\_min), and lower minimum eGFR values.


\subsection{Analysis of Model Misclassifications: Type I and Type II Errors}

\begin{figure}[htbp]
\begin{center}
\centering
\includegraphics[width=1\textwidth]{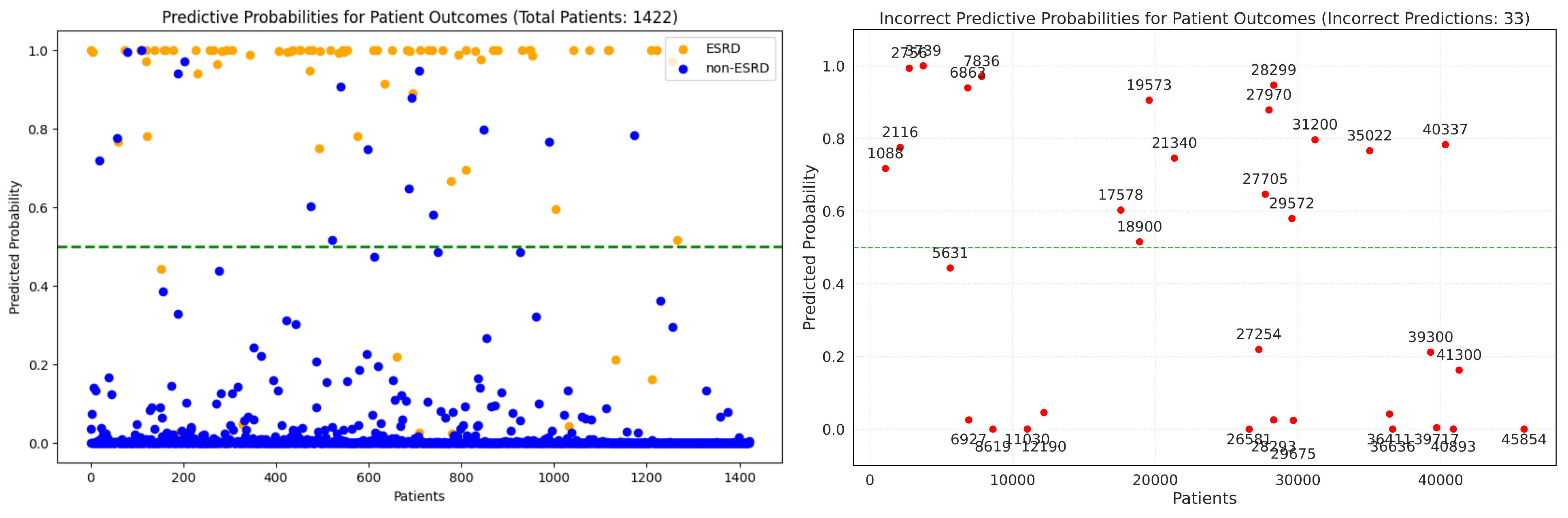}
\caption{Analysis of Model Misclassification: Type I and Type II Errors. The left panel shows predicted probabilities for all patient outcomes (24-month conversation window, n=1,422), stratified by ESRD (green) and non-ESRD (blue) cases, with patient indices (1-1422) on the x-axis. The right panel displays only the incorrectly predicted cases (n=33), using actual patient IDs on the x-axis for identification purposes. The horizontal dashed line at 0.5 represents the classification threshold.}
\label{fig:error_0}
\end{center}
\end{figure}

All subsequent analyses utilize our optimal 24-month observation window and the LSTM model, which demonstrated superior performance among all tested architectures. To explore these misclassifications, we analyzed Type I and Type II errors, as depicted in Figure~\ref{fig:error_0}. Notably, most incorrect predictions for patients who progressed to ESRD, yet were predicted otherwise, cluster near the lower end of the plot rather than around the 0.5 decision threshold or randomly scattered. This clustering suggests that consistent factors may be influencing these errors.


To understand model misclassification causes, we analyzed Type I and Type II errors. For Type II errors, we compared 16 false negatives (patients incorrectly predicted not to develop ESRD but who did) with 70 true positives (patients correctly predicted to progress). For Type I errors, we contrasted 17 false positives (patients incorrectly predicted to develop ESRD) with 1,319 true negatives (patients correctly predicted not to progress).

\vspace{.5cm}
\begin{table}[htbp]
  \caption{Analysis of Type I and Type II Errors (24-month observation window, n=1,422): Features and Their Impact on Prediction Accuracy. This table presents representative subsets of features under various timestamps. Features in bold appear in both analyses.}
  \vspace{.1cm}
  \centering
  
  \begin{subtable}[t]{\textwidth}
  \caption{Type II Error Analysis: Feature Comparison Between False Negatives and True Positives.}
  \vspace{-0.2cm}
    \centering
        \begin{tabular}{|c|cc|c|c|}
\hline
\multirow{2}{*}{Model} & \multicolumn{2}{c|}{Mean} & \multirow{2}{*}{P-value} & \multirow{2}{*}{Timestamp} \\ \cline{2-3}
                       & \multicolumn{1}{c|}{Correct} & Incorrect &                 &                            \\ \hline
n\_claims\_I           & \multicolumn{1}{c|}{0.70}    & 0.14      & 0.048752        & 5                          \\ \hline
\textbf{n\_claims\_O}  & \multicolumn{1}{c|}{5.71}    & 2.67      & 0.001325        & 2                          \\ \hline
net\_exp\_O            & \multicolumn{1}{c|}{3831.83} & 830.61    & 0.002747        & 0                          \\ \hline
\textbf{S5}           & \multicolumn{1}{c|}{0.36}    & 0.06      & 0.001189        & 6                          \\ \hline
\textbf{S5}           & \multicolumn{1}{c|}{0.37}    & 0.49      & 0.662073        & 7                          \\ \hline
\textbf{net\_exp\_O}   & \multicolumn{1}{c|}{4219.01} & 1044.37   & 0.001883        & 4                          \\ \hline
\end{tabular}
\vspace{0.5cm}
\label{table:error_2}
  \end{subtable}
  \vspace{0.5cm}  
  
  \begin{subtable}[t]{\textwidth}
    \caption{Type I Error Analysis: Feature Comparison Between False Positives and True Negatives. }
    \vspace{-0.2cm}
    \centering
        \begin{tabular}{|c|cc|c|c|}
        \hline
        \multirow{2}{*}{Model} & \multicolumn{2}{c|}{Mean} & \multirow{2}{*}{P-value} & \multirow{2}{*}{Timestamp} \\ \cline{2-3}
                               & \multicolumn{1}{c|}{Correct} & Incorrect &              &                            \\ \hline
        n\_claims\_DR          & \multicolumn{1}{c|}{13.52}   & 7.21      & 0.001156     & 3                          \\ \hline
        \textbf{n\_claims\_O}  & \multicolumn{1}{c|}{3.03}    & 1.69      & 0.009384     & 4                          \\ \hline
        S4                     & \multicolumn{1}{c|}{3831.83} & 830.61    & 0.002747     & 4                          \\ \hline
        \textbf{S5}           & \multicolumn{1}{c|}{0.09}    & 0.35      &  0.000250 & 6                          \\ \hline
        \textbf{S5}           & \multicolumn{1}{c|}{0.08}    & 0.35      &  0.001388 & 7                          \\ \hline
        \textbf{net\_exp\_O}   & \multicolumn{1}{c|}{1044.82} & 333.61    & 0.000324     & 4                          \\ \hline
        \end{tabular}
    \label{table:error_1}
\end{subtable}
  \vspace{0.5cm}  
  \label{table:error_table}
\end{table}



Based on Tables \ref{table:error_2} and \ref{table:error_1}, CKD Stage 5 emerges as a critical feature in prediction errors. For false negatives (incorrectly predicted as not progressing to ESRD), the mean value for CKD Stage 5 at timestamp 6 is significantly lower (0.06 versus 0.36) compared to correctly predicted cases, suggesting these patients didn't have a Stage 5 record until later. Conversely, false positives show CKD Stage 5 presence at both timestamps 6 and 7, indicating these patients had Stage 5 records but didn't actually progress to ESRD.

\subsection{Impact of Updated eGFR Equation on Racial Bias in Predictions}

\begin{figure}[htbp]
  \centering
  \begin{subfigure}[t]{0.49\textwidth}
    \centering
    \includegraphics[width=\textwidth]{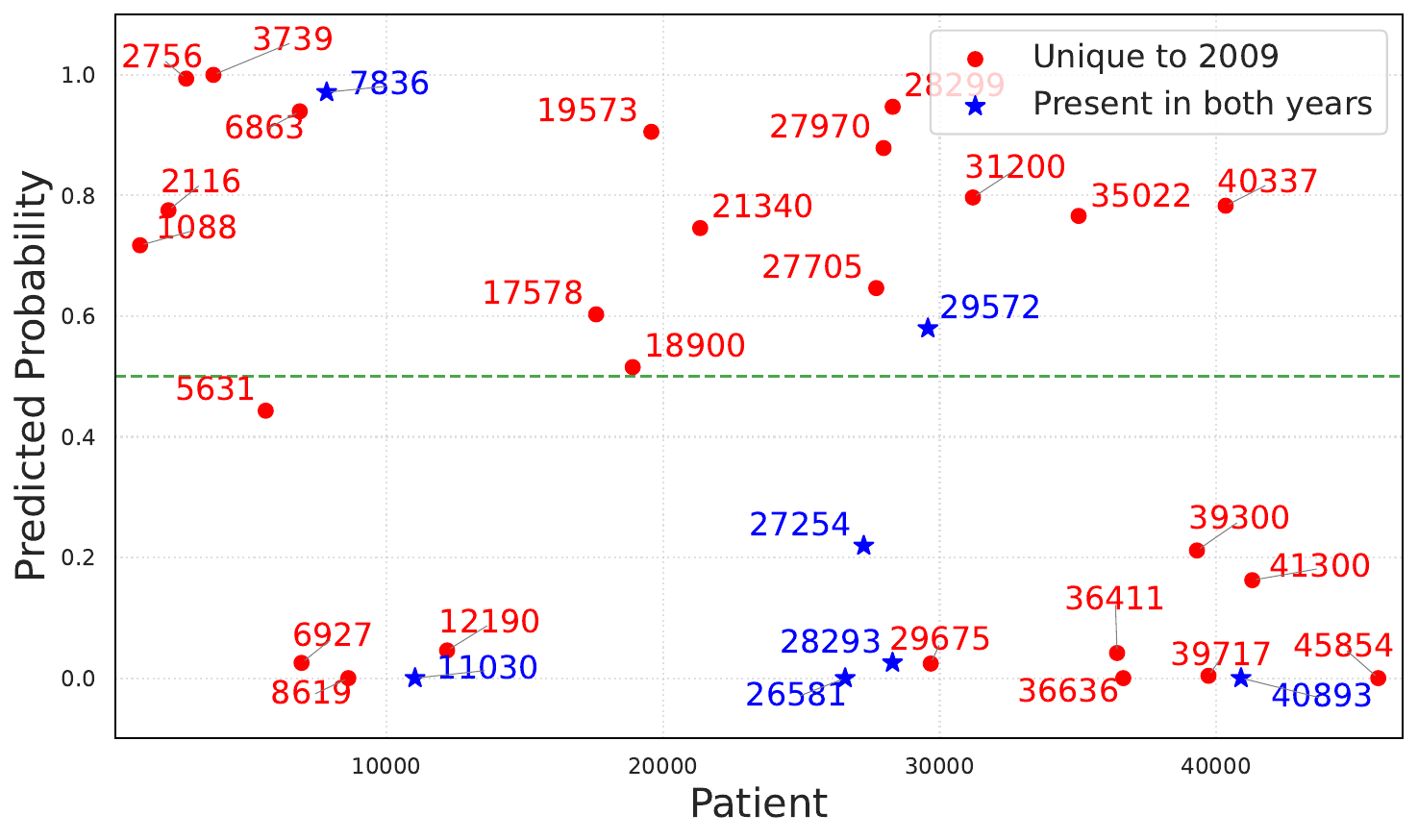}
    \subcaption{Incorrect predictive probabilities obtained using the 2009 eGFR equation.}
    \label{fig:last_1}
  \end{subfigure}
  \hfill
  \begin{subfigure}[t]{0.49\textwidth}
    \centering
    \includegraphics[width=\textwidth]{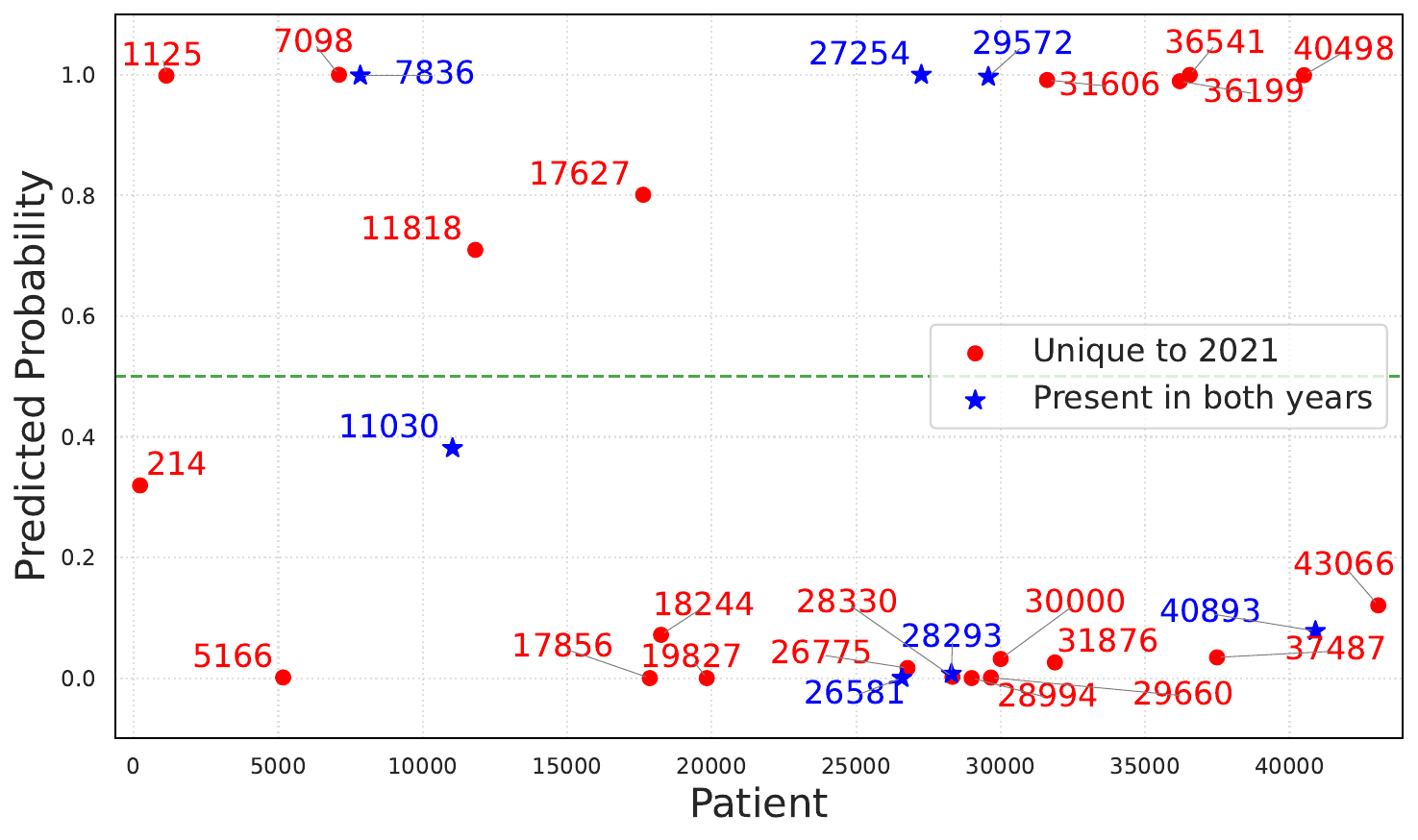}
    \subcaption{Incorrect predictive probabilities obtained using the 2021 eGFR equation.}
    \vspace{0.4cm}
    \label{fig:last_2}
  \end{subfigure}

  \caption{Impact of Updated eGFR Equations (2009 vs. 2021) on Prediction Errors. Subplots compare incorrect predictions using the two eGFR equations, highlighting the differences in misclassification patterns between the traditional and updated formulas.}
  \label{fig:2x2grid}
\end{figure}


When comparing predictions between the 2009 and 2021 eGFR equations (Fig.~\ref{fig:last_1}, ~\ref{fig:last_2}), we observed a decrease in total incorrect predictions from 33 to 28, reducing the overall error rate from 0.0232 to 0.0228. Among these misclassifications, only 7 cases were common to both equations. Analysis of racial distribution in prediction errors revealed varying patterns across different groups: incorrect predictions for white patients increased from 2 to 4, African American patients decreased from 4 to 1, and patients in the "Other" category increased from 10 to 12. However, these numbers are clearly too small for robust insights.


\section{Discussion}

This study demonstrates the value of integrating multiple data sources and deep learning methodologies for improving ESRD progression predictions in CKD patients. By combining claims and EHR data with LSTM and GRU models, we achieved enhanced predictive accuracy using a 24-month observation window—an optimal balance between early detection and prediction reliability. Our approach addresses a critical gap in chronic disease management by offering a more comprehensive view of patient health trajectories than single-source approaches.

Error analysis revealed important clinical implications. Patients incorrectly predicted not to develop ESRD (false negatives) had significantly lower prevalence of CKD Stage 5 records at timestamp 6 compared to true positives (0.06 vs 0.36), suggesting sudden kidney function decline—a phenomenon clinicians recognize as "crashing" into dialysis. Conversely, false positives showing CKD Stage 5 in consecutive timestamps without progression may reflect data censoring, as patients might have progressed after our dataset endpoint.

While the updated 2021 eGFR equation showed modest improvement in prediction accuracy compared to the 2009 equation (error rate reduction from 0.0232 to 0.0228), the pattern of misclassifications offers greater insight. With only 7 cases misclassified by both equations despite similar overall error rates, these formulas clearly capture different aspects of kidney function. This suggests that while the 2021 equation may better serve minoritized populations by removing race-based adjustments, eGFR alone remains insufficient for accurate ESRD prediction.

These findings underscore the need for a multi-faceted approach to risk assessment, combining traditional clinical markers, novel predictive features, and awareness of potential rapid disease progression patterns to develop more targeted interventions across diverse populations.

\subsection{Limitations}

This study’s reliance on data from a single institution may limit the model’s generalizability to other care settings. Using EHR data introduces observational bias, incomplete records, and underrepresentation of certain patient groups, which can undermine both accuracy and fairness. Although we applied oversampling to mitigate class imbalance, the contrast between high AUROC and lower F1/AUPRC indicates that imbalance remains an issue; more sophisticated approaches—such as ensemble methods, cost-sensitive learning, or hybrid sampling—may be needed. Finally, unaddressed time lags between claims and clinical data can distort temporal relationships and reduce predictive precision.

\subsection{Future Directions}

To address data censoring issues, we will first truncate data to 2016 and analyze outcomes from 2017-2018, then expand our dataset beyond 2018 to enhance trajectory modeling. We'll integrate unstructured clinical notes to capture patient information missed in structured data, providing a more comprehensive health view and improving prediction accuracy.

We will implement advanced algorithms to synchronize claims and clinical data temporally, reducing time lag effects that currently impact model performance. To understand prediction errors, we'll perform SHAP analysis on misclassified cases, identifying key features contributing to these misclassifications and guiding targeted model improvements.

Finally, we'll validate our framework's versatility by applying it to other chronic conditions such as heart disease, assessing its broader potential across various care delivery settings. This systematic expansion will provide actionable insights for both our ESRD prediction model and chronic disease management more broadly.


\section{Conclusion}

This study demonstrates the effectiveness of integrating diverse healthcare data with advanced machine learning techniques to accurately predict ESRD in CKD patients. The combined-data approach substantially enhances predictive performance and provides deeper insights into disease progression. SHAP analysis and feature importance assessment highlighted key predictors at both individual and cohort levels.

A critical contribution of this work is a framework for optimizing observation windows, balancing early detection with prediction accuracy. We identified a 24-month observation window as optimal for maximizing predictive effectiveness while minimizing unnecessary interventions. Additionally, evaluating the updated 2021 eGFR equation supports efforts towards health equity and fairer clinical outcomes.

Overall, this research advances CKD management through integrated data and innovative AI methodologies, setting the stage for personalized, equitable care and future applications in chronic disease prediction and management.


\section{Acknowledgments}

The authors sincerely thank the community nephrology practice for providing access to their de-identified clinical data and the wealth of medical knowledge and inputs shared by several nephrologists in the practice and their CIO on the raw data and the results of our analysis. Similarly, we are very grateful to the health insurance organization for providing access to the de-identified claims data on their shared set of patients with the nephrology practice and for the invaluable help provided by their data team in understanding the claims data.

\section*{Funding Statement}

This work was supported by the fellowship support for Y.L. from the Center for Machine Learning and Health (CMLH) at Carnegie Mellon University. 

\section*{Competing Interests Statement}  

The authors have no competing interests to declare.

\section*{Contributorship Statement}  

Both authors meet the International Committee of Medical Journal Editors (ICMJE) authorship criteria. Rema supervised the research, advised on study conception and design, reviewed and critically revised the manuscript. Yubo contributed to study conception and design, implemented all computational methods, wrote the manuscript draft, and performed data analysis. Both authors approved the final manuscript version and agreed to be accountable for all aspects of the work, ensuring its accuracy and integrity.

AI Writing Assistance: All drafts of this manuscript were created by the authors without AI assistance. GPT-4.5 and Claude 3.7 (Sonnet) were used solely for polishing grammar and refining expression. The outputs from these tools were critically compared and reviewed by the authors to ensure accuracy and clarity. Final revisions were made by the authors to ensure all expressions align with the intended meaning and academic standards.

\section*{Data Availability Statement}

The data underlying this article consist of active healthcare clinical and claims data, which cannot be shared publicly due to privacy regulations and the sensitive nature of the information.

\newpage
\renewcommand{\bibsection}{\centering\section*{\refname}}
\makeatletter
\renewcommand{\@biblabel}[1]{\hfill #1.}
\makeatother

\bibliographystyle{vancouver}
\bibliography{amia}  

\begin{thebibliography}{10}

\bibitem{NKF_CKD_2024}
Foundation NK.
\newblock Chronic Kidney Disease (CKD).
\newblock National Kidney Foundation. 2024.
\newblock Available from: \url{https://www.kidney.org/atoz/content/about-chronic-kidney-disease}.

\bibitem{lin2013progression}
Lin CM, Yang MC, Hwang SJ, Sung JM.
\newblock Progression of stages 3b--5 chronic kidney disease—preliminary results of Taiwan National pre-ESRD disease management program in Southern Taiwan.
\newblock Journal of the Formosan Medical Association. 2013;112(12):773-82.

\bibitem{NCHS2019Mortality}
for Health~Statistics NC.
\newblock Mortality in the United States.
\newblock https://wwwcdcgov/nchs/products/databriefs/db395htm. 2019.

\bibitem{guo2020machine}
Guo Y, Yu H, Chen D, Zhao YY.
\newblock Machine learning distilled metabolite biomarkers for early stage renal injury.
\newblock Metabolomics. 2020;16(1):1-10.

\bibitem{belur2020machine}
Belur~Nagaraj S, Pena MJ, Ju W, Heerspink HL, Consortium BD.
\newblock Machine-learning--based early prediction of end-stage renal disease in patients with diabetic kidney disease using clinical trials data.
\newblock Diabetes, Obesity and Metabolism. 2020;22(12):2479-86.

\bibitem{krishnamurthy2021machine}
Krishnamurthy V, Gopal A, Kannan R, et~al.
\newblock Machine learning for predicting outcomes in chronic kidney disease: Challenges and opportunities.
\newblock IEEE Journal of Biomedical and Health Informatics. 2021;25(4):1319-31.

\bibitem{li2024towards}
Li Y, Al-Sayouri S, Padman R.
\newblock Towards Interpretable End-Stage Renal Disease (ESRD) Prediction: Utilizing Administrative Claims Data with Explainable AI Techniques.
\newblock In: AMIA Annual Symposium Proceedings. American Medical Informatics Association; 2024. .

\bibitem{sharma2020model}
Sharma A, Alvarez PJ, Woods SD, Dai D.
\newblock A Model to Predict Risk of Hyperkalemia in Patients with Chronic Kidney Disease Using a Large Administrative Claims Database.
\newblock ClinicoEconomics and Outcomes Research: CEOR. 2020;12:657.

\bibitem{tangri2011predictive}
Tangri N, Stevens LA, Griffith J, Tighiouart H, Djurdjev O, Naimark D, et~al.
\newblock A predictive model for progression of chronic kidney disease to kidney failure.
\newblock JAMA. 2011;305(15):1553-9.

\bibitem{NKF_eGFR}
Foundation NK, {}. Estimated Glomerular Filtration Rate (eGFR). National Kidney Foundation; n.d.
\newblock Accessed: 2024-12-16.
\newblock \url{https://www.kidney.org/kidney-topics/estimated-glomerular-filtration-rate-egfr}.

\bibitem{sun2020development}
Sun L, Shang J, Xiao J, Zhao Z.
\newblock Development and validation of a predictive model for end-stage renal disease risk in patients with diabetic nephropathy confirmed by renal biopsy.
\newblock PeerJ. 2020;8:e8499.

\bibitem{vanRijsbergen1979}
van Rijsbergen CJ.
\newblock Information Retrieval.
\newblock 2nd ed. Butterworth-Heinemann; 1979.
\newblock Introduced the F-measure, which includes the F1-score.

\bibitem{hanley1982meaning}
Hanley JA, McNeil BJ.
\newblock The meaning and use of the area under a receiver operating characteristic (ROC) curve.
\newblock Radiology. 1982;143(1):29-36.

\bibitem{davis2006pr}
Davis J, Goadrich M.
\newblock The relationship between Precision-Recall and ROC curves.
\newblock In: Proceedings of the 23rd International Conference on Machine Learning (ICML). ACM; 2006. p. 233-40.

\bibitem{white2011multiple}
White IR, Royston P, Wood AM.
\newblock Multiple imputation using chained equations: issues and guidance for practice.
\newblock Statistics in medicine. 2011;30(4):377-99.

\bibitem{sedgwick2012log}
Sedgwick P.
\newblock Log transformation of data.
\newblock BMJ. 2012;345.

\bibitem{aryal2018anomaly}
Aryal S.
\newblock Anomaly Detection Technique Robust to Units and Scales of Measurement.
\newblock In: Advances in Knowledge Discovery and Data Mining (PAKDD 2018). Springer; 2018. p. 589-601.
\newblock Available from: \url{https://link.springer.com/chapter/10.1007/978-3-319-93034-3_47}.

\bibitem{NIDDK_UACR}
Foundation NK, {}. Urine Albumin-to-Creatinine Ratio (UACR). National Kidney Foundation; n.d.
\newblock Accessed: 2024-12-16.
\newblock \url{https://www.kidney.org/kidney-topics/urine-albumin-creatinine-ratio}.

\bibitem{chawla2002smote}
Chawla NV, Bowyer KW, Hall LO, Kegelmeyer WP.
\newblock SMOTE: Synthetic Minority Over-sampling Technique.
\newblock Journal of Artificial Intelligence Research. 2002;16:321-57.
\newblock Available from: \url{https://www.jair.org/index.php/jair/article/view/10302}.

\bibitem{saif2024early}
Saif D, Sarhan AM, Elshennawy NM.
\newblock Early prediction of chronic kidney disease based on ensemble of deep learning models and optimizers.
\newblock Journal of electrical systems and information technology. 2024;11(1):17.

\bibitem{kleinbaum2010logistic}
Kleinbaum DG, Klein M.
\newblock Logistic Regression: A Self-Learning Text.
\newblock 3rd ed. Springer Science \& Business Media; 2010.

\bibitem{breiman2001random}
Breiman L.
\newblock Random Forests.
\newblock Machine Learning. 2001;45(1):5-32.

\bibitem{chen2016xgboost}
Chen T, Guestrin C.
\newblock XGBoost: A Scalable Tree Boosting System.
\newblock Proceedings of the 22nd ACM SIGKDD International Conference on Knowledge Discovery and Data Mining. 2016:785-94.

\bibitem{wong2019reliable}
Wong TT, Yeh PY.
\newblock Reliable accuracy estimates from k-fold cross validation.
\newblock IEEE Transactions on Knowledge and Data Engineering. 2019;32(8):1586-94.

\bibitem{burckhardt}
Burckhardt P, Nagin D, Padman R.
\newblock Multi-Trajectory Models of Chronic Kidney Disease Progression.
\newblock AMIA Annu Symp Proc. 2017:1737-46.

\bibitem{lecun2015deep}
LeCun Y, Bengio Y, Hinton G.
\newblock Deep learning.
\newblock Nature. 2015;521(7553):436-44.

\bibitem{rumelhart1986learning}
Rumelhart DE, Hinton GE, Williams RJ.
\newblock Learning representations by back-propagating errors.
\newblock Nature. 1986;323(6088):533-6.

\bibitem{hochreiter1997long}
Hochreiter S, Schmidhuber J.
\newblock Long short-term memory.
\newblock Neural computation. 1997;9(8):1735-80.

\bibitem{cho2014learning}
Cho K, Van~Merri{\"e}nboer B, Gulcehre C, Bahdanau D, Bougares F, Schwenk H, et~al.
\newblock Learning phrase representations using RNN encoder-decoder for statistical machine translation.
\newblock In: Proceedings of the 2014 Conference on Empirical Methods in Natural Language Processing (EMNLP). Association for Computational Linguistics; 2014. p. 1724-34.

\bibitem{bai2018empirical}
Bai S, Kolter JZ, Koltun V.
\newblock An empirical evaluation of generic convolutional and recurrent networks for sequence modeling.
\newblock arXiv preprint arXiv:180301271. 2018.

\bibitem{inker2021new}
Inker LA, Eneanya ND, Coresh J, Tighiouart H, Wang D, Sang Y, et~al.
\newblock New creatinine-and cystatin C--based equations to estimate GFR without race.
\newblock New England Journal of Medicine. 2021;385(19):1737-49.

\bibitem{miller2022national}
Miller WG, Kaufman HW, Levey AS, Straseski JA, Wilhelms KW, Yu HY, et~al.
\newblock National Kidney Foundation Laboratory Engagement Working Group recommendations for implementing the CKD-EPI 2021 race-free equations for estimated glomerular filtration rate: practical guidance for clinical laboratories.
\newblock Clinical chemistry. 2022;68(4):511-20.

\bibitem{levey2020kidney}
Levey AS, Titan SM, Powe NR, Coresh J, Inker LA.
\newblock Kidney disease, race, and GFR estimation.
\newblock Clinical Journal of the American Society of Nephrology. 2020;15(8):1203-12.

\bibitem{diao2021clinical}
Diao JA, Inker LA, Levey AS, Tighiouart H, Powe NR, Crews DC, et~al.
\newblock Clinical implications of removing race from estimates of kidney function.
\newblock JAMA. 2021;325(2):184-96.

\end{thebibliography}

\newpage
\section*{Supplementary}
\appendix
\beginsupplement
\section{Detailed Data Source Information}
\label{app:data_details}

\subsection*{Administrative Claims Data}
The de-identified administrative claims dataset used in this study was provided by Highmark, a major health insurance provider serving Pennsylvania, Delaware, and West Virginia. The dataset includes comprehensive records of patient interactions with healthcare professionals from January 1, 2009, to December 31, 2018. The use of this dataset was approved by Highmark's Institutional Review Board.

\subsection*{Clinical EHR Data}
The clinical dataset was sourced from the Electronic Health Record (EHR) data of The Nephrology Associates (TMA, P.C.), a leading community nephrology practice serving the greater Pittsburgh region. This dataset contains detailed laboratory results, patient demographics, diagnostic information, and medication records spanning from 1998 to 2018. For consistency of integration with the claims data, we truncated the clinical data to match the 10-year span of the claims data (2009–2018). The usage of the clinical data was approved by the Institutional Review Board of Carnegie Mellon University.

\section{Laboratory Test Value Ranges for Distribution Detection and Adjustment}
\label{app:ranges}
\begin{table}[H]
  \caption{The table below presents the reference ranges established for laboratory tests during the data preprocessing phase. These ranges were used for distribution detection and adjustment to identify and correct unit inconsistencies.}
  \centering

\begin{tabular}{|l|l|l|l|}
\hline Laboratory Test & Minimum Value & Maximum Value & Unit Conversion Notes \\
\hline Serum Creatinine & 0 & 15 & $1 \mathrm{mg} / \mathrm{dL}=88.4 \mu \mathrm{~mol} / \mathrm{L}$ \\
\hline Intact PTH & 0 & 500 & $50 \mathrm{pg} / \mathrm{mL}=5.263 \mathrm{pmol} / \mathrm{L}$ \\
\hline Hemoglobin & 2 & 20 & $1 \mathrm{~g} / \mathrm{dL}=10 \mathrm{~g} / \mathrm{L} ; 1 \mathrm{~g} / \mathrm{dL}=100 / 645$ $\mathrm{mmol} / \mathrm{L}$ \\
\hline Albumin to Creatinine Ratio, Urine & 0 & 600 & $1 \mathrm{mg} / \mathrm{g}=1000 \mathrm{mg} / \mathrm{mg}$ \\
\hline Bicarbonate (Mostly Arterial) & 0 & 30 & $\mathrm{mEq} / \mathrm{L}=\mathrm{mmol} / \mathrm{L}$ \\
\hline Phosphorus & 0 & 10 & $4 \mathrm{mg} / \mathrm{dL}=0.00129 \mathrm{mmol} / \mathrm{L}$ \\
\hline Serum Calcium & 0 & 20 & $10 \mathrm{mg} / \mathrm{dL}=0.00250 \mathrm{mmol} / \mathrm{L}$ \\
\hline Urine Albumin & 0 & 600 & - \\
\hline eGFR & 0 & 150 & - \\
\hline
\end{tabular}
  \label{tab:ranges}
\end{table}

\newpage
\section{Comorbidity Feature ICD Code Prefixes}
\label{app:icd}
\begin{table}[ht]
  \centering
  \caption{Comorbidity features with ICD-9 and ICD-10 code prefixes used for feature extraction in this study}
  \label{tab:comorbidity-code-prefixes}
  \begin{tabular}{|l|p{0.3\textwidth}|p{0.3\textwidth}|}
    \hline
    \textbf{Condition}                    & \textbf{ICD-9 Prefixes}   & \textbf{ICD-10 Prefixes}            \\ 
    \hline
    Diabetes                              & 249*, 250*                & E08*–E13*                           \\ \hline
    Hypertension                          & 401*–405*                 & I10*–I15*                           \\ \hline
    Cardiovascular Disease                & 429*                      & I25*                                \\ \hline
    Anemia                                & 285.9                     & D64.9, D50*–D59*                    \\ \hline
    Metabolic Acidosis                    & 276.2                     & E87.2                               \\ \hline
    Proteinuria                           & 791.0                     & R80.9                               \\ \hline
    Secondary Hyperparathyroidism         & 252.02, 588.81            & E21.1, N25.81                       \\ \hline
    Phosphatemia                          & 275.3                     & E83.39                              \\ \hline
    Atherosclerosis                       & 440*                      & I70*                                \\ \hline
    Congestive Heart Failure (CHF)        & 428*                      & I50*                                \\ \hline
    Conduction \& Dysrhythmias            & 426*–427*                 & I44*–I49*, Z450*, Z958*             \\ \hline
    Myocardial Infarction (MI)            & 410*                      & I21*                                \\ \hline
    CVD Other                             & 390*–459*                 & I05*–I09*, I30*–I52*                \\ \hline
    Stroke                                & 433*–434*                 & I63*                                \\ \hline
    Fluid/Electrolytes                    & 276*                      & E86*–E87*                           \\ \hline
    Mineral Disorders                     & 275*                      & E20*–E21*, E83*, N25*               \\ \hline
    Nutritional Deficiencies              & 260*–269*                 & E40*–E46*, E50*–E64*, D50*          \\ \hline
    CKD Stage 4                           & 585.4*                    & N18.4*                              \\ \hline
    CKD Stage 5                           & 585.5*                    & N18.5*                              \\ \hline
    CKD End-Stage Renal Disease (ESRD)    & 585.6*                    & N18.6*                              \\ \hline
  \end{tabular}
\end{table}

Note: prefix * matches any sequence of trailing digits. For example, I44*–I49* captures all codes from I440 through I499 such as I4402 and I4439; Z450* captures Z4502, Z45010, Z45018, etc.

\newpage
\section{Neural Network Implementation Details}
\label{app:implementation_details}

\subsection*{Hardware Specifications}
\begin{table}[ht]
\centering
\begin{tabular}{|l|l|}
\hline
\textbf{Resource} & \textbf{Specification} \\
\hline
GPU & 2 $\times$ NVIDIA RTX 4090 (24GB VRAM) \\
\hline
RAM & 512GB DDR4 \\
\hline
Storage & 2TB NVMe SSD \\
\hline
\end{tabular}
\caption{Hardware specifications used for model training and evaluation.}
\label{tab:hardware}
\end{table}

\subsection*{Software Environment and Library Versions}
\begin{table}[ht]
\centering
\begin{tabular}{|l|l|l|}
\hline
\textbf{Component} & \textbf{Version} & \textbf{Purpose} \\
\hline
Python & 3.10.8 & Programming language \\
\hline
CUDA & 11.6 & GPU acceleration library \\
\hline
PyTorch & 1.13.0 & Deep learning framework \\
\hline
H2O & 3.38.0.1 & Machine learning model tuning \\
\hline
NNI & 2.10 & Neural architecture search \\
\hline
scikit-learn & 1.1.3 & Traditional ML algorithms \\
\hline
pandas & 1.5.1 & Data manipulation and analysis \\
\hline
NumPy & 1.23.4 & Numerical computing \\
\hline
\end{tabular}
\caption{Software environment and library versions used for model implementation.}
\label{tab:software}
\end{table}

\subsection*{Model Architecture, Hyperparameters, and Training Details}
\begin{table}[h!]
\centering
\small
\begin{tabular}{|l|p{4.5cm}|c|p{5.5cm}|c|}\hline
\textbf{Model} & \textbf{Architecture Details} & \textbf{Parameters} & \textbf{Hyperparameters} & \textbf{Training Time} \\
\hline
CNN & \begin{tabular}[c]{@{}l@{}}3 conv layers (64, 128, 256 filters),\\ 2 FC layers (512, 256)\end{tabular} & 4.3M & 
\begin{tabular}[c]{@{}l@{}}Batch size: 32, Optimizer: Adam, \\ Learning rate: 1e-4, Dropout: 0.3\end{tabular} & 8.5 hours \\
\hline
RNN & \begin{tabular}[c]{@{}l@{}}2 recurrent layers (128 units),\\ 2 FC layers (256, 128)\end{tabular} & 2.1M & 
\begin{tabular}[c]{@{}l@{}}Batch size: 64, Optimizer: RMSprop\\ Learning rate: 5e-4, Dropout: 0.2\end{tabular} & 9.2 hours \\
\hline
LSTM & \begin{tabular}[c]{@{}l@{}}2 LSTM layers (256 units),\\ 1 FC layer (128)\end{tabular} & 5.7M & 
\begin{tabular}[c]{@{}l@{}}Batch size: 32, Optimizer: Adam\\ Learning rate: 1e-4, Dropout: 0.3\end{tabular} & 15.3 hours \\
\hline
GRU & \begin{tabular}[c]{@{}l@{}}2 GRU layers (256 units),\\ 1 FC layer (128)\end{tabular} & 4.9M & 
\begin{tabular}[c]{@{}l@{}}Batch size: 32, Optimizer: Adam\\ Learning rate: 1e-4, Dropout: 0.3\end{tabular} & 13.7 hours \\
\hline
TCN & \begin{tabular}[c]{@{}l@{}}4 temporal blocks,\\ kernel size 3, 128 filters\end{tabular} & 3.8M & 
\begin{tabular}[c]{@{}l@{}}Batch size: 32, Optimizer: Adam\\ Learning rate: 2e-4, Dropout: 0.2\end{tabular} & 12.1 hours \\
\hline
\end{tabular}
\caption{Model architecture, hyperparameters, and training details for each neural network model.}
\label{tab:models}
\end{table}

\newpage
\section{Model Performance Metrics Across Observation Windows}
\label{app:performance}
\begin{table}[H]
  \caption{Comparison of AUROC and F1 scores across different models, using claims data-only, clinical data-only, and merged data (24-month observation window, n=1,422). Bold font is used to mark the best performance within each category (ML or DL methods), while bold and italic font highlights the best overall performance across all methods (ML + DL).}
  \centering
  \begin{subtable}[t]{\textwidth}
  \caption{Performance of models using claims data only.}

    \centering

    \begin{tabular}{|cc|c|cc|cc|cc|cc|cc|}
    \hline
    \multicolumn{2}{|c|}{\multirow{2}{*}{}}                                                                                                                                                                          &                     & \multicolumn{2}{c|}{\textbf{6 months}}    & \multicolumn{2}{c|}{\textbf{12 months}}   & \multicolumn{2}{c|}{\textbf{18 months}}   & \multicolumn{2}{c|}{\textbf{24 months}}   & \multicolumn{2}{c|}{\textbf{30 months}}   \\ \cline{3-13} 
    \multicolumn{2}{|c|}{}                                                                                                                                                                                           & Model/Metrics             & \multicolumn{1}{c|}{AUC}  & F1   & \multicolumn{1}{c|}{AUC}  & F1   & \multicolumn{1}{c|}{AUC}  & F1   & \multicolumn{1}{c|}{AUC}  & F1   & \multicolumn{1}{c|}{AUC}  & F1   \\ \hline
    \multicolumn{1}{|c|}{\multirow{8}{*}{\begin{tabular}[c]{@{}c@{}}Claims\\ Data\\ Only  \\ Modeling\end{tabular}}} & \multirow{3}{*}{\begin{tabular}[c]{@{}c@{}}Machine \\ Learning \\ Methods\end{tabular}}         & Logistic Regression & \multicolumn{1}{c|}{0.61} & 0.24 & \multicolumn{1}{c|}{0.64} & 0.31 & \multicolumn{1}{c|}{0.70} & 0.32 & \multicolumn{1}{c|}{0.72} & 0.33 & \multicolumn{1}{c|}{0.71} & 0.29 \\ \cline{3-13} 
\multicolumn{1}{|c|}{}                                                                                           &                                                                                                     & Random Forest       & \multicolumn{1}{c|}{0.69} & 0.26 & \multicolumn{1}{c|}{0.73} & 0.33 & \multicolumn{1}{c|}{0.74} & 0.41 & \multicolumn{1}{c|}{0.74} & 0.36 & \multicolumn{1}{c|}{0.68} & 0.33 \\ \cline{3-13} 
\multicolumn{1}{|c|}{}                                                                                           &                                                                                                     & XGBoost             & \multicolumn{1}{c|}{0.70} & 0.27 & \multicolumn{1}{c|}{0.75} & 0.35 & \multicolumn{1}{c|}{\textbf{0.78}} & \textbf{0.42} & \multicolumn{1}{c|}{0.75} & 0.39 & \multicolumn{1}{c|}{0.69} & 0.38 \\ \cline{2-13} 
    \multicolumn{1}{|c|}{}                                                                                         & \multirow{5}{*}{\begin{tabular}[c]{@{}c@{}}Deep\\ Learning\\ Methods\end{tabular}} &  CNN                 & \multicolumn{1}{c|}{0.72} & 0.36 & \multicolumn{1}{c|}{0.73} & 0.42 & \multicolumn{1}{c|}{0.81} & 0.44 & \multicolumn{1}{c|}{0.82} & 0.45 & \multicolumn{1}{c|}{0.84} & 0.44 \\ \cline{3-13} 
\multicolumn{1}{|c|}{}                                                                                           &                                                                                                     & RNN                 & \multicolumn{1}{c|}{0.77} & 0.39 & \multicolumn{1}{c|}{0.75} & 0.44 & \multicolumn{1}{c|}{0.83} & 0.49 & \multicolumn{1}{c|}{0.90} & 0.50 & \multicolumn{1}{c|}{0.82} & 0.47 \\ \cline{3-13} 
\multicolumn{1}{|c|}{}                                                                                           &                                                                                                     & LSTM                & \multicolumn{1}{c|}{0.76} & 0.33 & \multicolumn{1}{c|}{0.79} & 0.43 & \multicolumn{1}{c|}{0.86} & 0.45 & \multicolumn{1}{c|}{\textit{\textbf{0.92}}} & {\textit{\textbf{0.54}}} & \multicolumn{1}{c|}{0.87} & 0.44 \\ \cline{3-13} 
\multicolumn{1}{|c|}{}                                                                                           &                                                                                                     & GRU                 & \multicolumn{1}{c|}{0.70} & 0.34 & \multicolumn{1}{c|}{0.78} & 0.46 & \multicolumn{1}{c|}{0.88} & 0.45 & \multicolumn{1}{c|}{0.91} & 0.50 & \multicolumn{1}{c|}{0.86} & 0.42 \\ \cline{3-13} 
\multicolumn{1}{|c|}{}                                                                                           &                                                                                                     & TCN                 & \multicolumn{1}{c|}{0.72} & 0.36 & \multicolumn{1}{c|}{0.81} & 0.37 & \multicolumn{1}{c|}{0.85} & 0.46 & \multicolumn{1}{c|}{0.88} & 0.52 & \multicolumn{1}{c|}{0.85} & 0.43 \\ \hline
    \end{tabular}
    \vspace{0.5cm}
    
    \label{table:case_1_1}
  \end{subtable}
  \vspace{0.5cm}  
  \begin{subtable}[t]{\textwidth}
\caption{Performance of models using clinical data only.}
    \centering

    \begin{tabular}{|cc|c|ll|ll|ll|ll|ll|}
    \hline
    \multicolumn{2}{|c|}{\multirow{2}{*}{}}                                                                                                                                                                     &                     & \multicolumn{2}{c|}{\textbf{6 months}}                       & \multicolumn{2}{c|}{\textbf{12 months}}                      & \multicolumn{2}{c|}{\textbf{18 months}}                      & \multicolumn{2}{c|}{\textbf{24 months}}                      & \multicolumn{2}{c|}{\textbf{30 months}}                      \\ \cline{3-13} 
    \multicolumn{2}{|c|}{}                                                                                                                                                                                      & Model/Metrics              & \multicolumn{1}{c|}{AUC}  & \multicolumn{1}{c|}{F1} & \multicolumn{1}{c|}{AUC}  & \multicolumn{1}{c|}{F1} & \multicolumn{1}{c|}{AUC}  & \multicolumn{1}{c|}{F1} & \multicolumn{1}{c|}{AUC}  & \multicolumn{1}{c|}{F1} & \multicolumn{1}{c|}{AUC}  & \multicolumn{1}{c|}{F1} \\ \hline
    \multicolumn{1}{|c|}{\multirow{8}{*}{\begin{tabular}[c]{@{}c@{}}Clinical\\ Data\\ Only   \\ Modeling\end{tabular}}} & \multirow{3}{*}{\begin{tabular}[c]{@{}c@{}}Machine\\ Learning\\   Methods\end{tabular}} & Logistic Regression & \multicolumn{1}{l|}{0.63} & 0.46                    & \multicolumn{1}{l|}{0.70} & 0.49                    & \multicolumn{1}{l|}{0.76} & 0.51                    & \multicolumn{1}{l|}{0.76} & 0.54                    & \multicolumn{1}{l|}{0.75} & 0.53                    \\ \cline{3-13} 
\multicolumn{1}{|c|}{}                                                                                             &                                                                                                     & Random Forest       & \multicolumn{1}{l|}{0.70} & 0.53                    & \multicolumn{1}{l|}{0.76} & 0.53                    & \multicolumn{1}{l|}{\textbf{0.80}} & 0.58                 & \multicolumn{1}{l|}{0.79} & 0.58                    & \multicolumn{1}{l|}{0.77} & 0.54                    \\ \cline{3-13} 
\multicolumn{1}{|c|}{}                                                                                             &                                                                                                     & XGBoost             & \multicolumn{1}{l|}{0.73} & 0.52                    & \multicolumn{1}{l|}{0.76} & 0.55                    & \multicolumn{1}{l|}{\textbf{0.80}} & \textbf{0.59 }                   & \multicolumn{1}{l|}{\textbf{0.80}} & 0.57                    & \multicolumn{1}{l|}{0.72} & 0.55                    \\ \cline{2-13} 
    \multicolumn{1}{|c|}{}                                                                                            & \multirow{5}{*}{\begin{tabular}[c]{@{}c@{}}Deep\\ Learning\\ Methods\end{tabular}}      & CNN                 & \multicolumn{1}{l|}{0.73} & 0.51                    & \multicolumn{1}{l|}{0.75} & 0.54                    & \multicolumn{1}{l|}{0.80} & 0.55                    & \multicolumn{1}{l|}{0.84} & 0.56                    & \multicolumn{1}{l|}{0.80} & 0.52                    \\ \cline{3-13} 
\multicolumn{1}{|c|}{}                                                                                             &                                                                                                     & RNN                 & \multicolumn{1}{l|}{0.74} & 0.53                    & \multicolumn{1}{l|}{0.77} & 0.57                    & \multicolumn{1}{l|}{0.82} & 0.60                    & \multicolumn{1}{l|}{0.85} & 0.61                    & \multicolumn{1}{l|}{0.79} & 0.55                    \\ \cline{3-13} 
\multicolumn{1}{|c|}{}                                                                                             &                                                                                                     & LSTM                & \multicolumn{1}{l|}{0.77} & 0.54                    & \multicolumn{1}{l|}{0.78} & 0.56                    & \multicolumn{1}{l|}{\textit{\textbf{0.84}}} & 0.63                    & \multicolumn{1}{l|}{\textit{\textbf{0.88}}} & 0.60                    & \multicolumn{1}{l|}{0.82} & 0.57                    \\ \cline{3-13} 
\multicolumn{1}{|c|}{}                                                                                             &                                                                                                     & GRU                 & \multicolumn{1}{l|}{0.80} & 0.54                    & \multicolumn{1}{l|}{0.77} & 0.57                    & \multicolumn{1}{l|}{0.83} & 0.61                    & \multicolumn{1}{l|}{0.87} & 0.60                    & \multicolumn{1}{l|}{0.84} & 0.53                    \\ \cline{3-13} 
\multicolumn{1}{|c|}{}                                                                                             &                                                                                                     & TCN                 & \multicolumn{1}{l|}{0.72} & 0.50                    & \multicolumn{1}{l|}{0.73} & 0.53                    & \multicolumn{1}{l|}{0.80} & 0.57                    & \multicolumn{1}{l|}{0.83} & 0.61                    & \multicolumn{1}{l|}{0.81} & 0.55                    \\ \hline
    \end{tabular}

    \label{table:case_1_2}
  \end{subtable}
  \vspace{0.5cm}  
  \begin{subtable}[t]{\textwidth}
    \caption{Performance of models using merged data.}
    \centering

\begin{tabular}{|cc|c|cc|cc|cc|cc|cc|}
\hline
\multicolumn{2}{|c|}{\multirow{2}{*}{}}                                                                                                                                                                           &                     & \multicolumn{2}{c|}{\textbf{6 months}}    & \multicolumn{2}{c|}{\textbf{12 months}}   & \multicolumn{2}{c|}{\textbf{18 months}}   & \multicolumn{2}{c|}{\textbf{24 months}}   & \multicolumn{2}{c|}{\textbf{30 months}}   \\ \cline{3-13} 
\multicolumn{2}{|c|}{}                                                                                                                                                                                            & Model/Metrics              & \multicolumn{1}{c|}{AUC}  & F1   & \multicolumn{1}{c|}{AUC}  & F1   & \multicolumn{1}{c|}{AUC}  & F1   & \multicolumn{1}{c|}{AUC}  & F1   & \multicolumn{1}{c|}{AUC}  & F1   \\ \hline
\multicolumn{1}{|c|}{\multirow{8}{*}{\begin{tabular}[c]{@{}c@{}}Merged\\ Data\\ Modeling\end{tabular}}} & \multirow{3}{*}{\begin{tabular}[c]{@{}c@{}}Machine \\ Learning \\ Methods\end{tabular}} & Logistic Regression & \multicolumn{1}{c|}{0.65} & 0.46 & \multicolumn{1}{c|}{0.69} & 0.47 & \multicolumn{1}{c|}{0.76} & 0.54 & \multicolumn{1}{c|}{0.75} & 0.55 & \multicolumn{1}{c|}{0.78} & 0.54 \\ \cline{3-13} 
\multicolumn{1}{|c|}{}                                                                                      &                                                                                                     & Random Forest       & \multicolumn{1}{c|}{0.74} & 0.44 & \multicolumn{1}{c|}{0.77} & 0.52 & \multicolumn{1}{c|}{0.83} & 0.58 & \multicolumn{1}{c|}{0.84} & 0.60 & \multicolumn{1}{c|}{0.80} & 0.55 \\ \cline{3-13} 
\multicolumn{1}{|c|}{}                                                                                      &                                                                                                     & XGBoost             & \multicolumn{1}{c|}{0.73} & 0.45 & \multicolumn{1}{c|}{0.76} & 0.50 & \multicolumn{1}{c|}{0.82} & 0.58 & \multicolumn{1}{c|}{\textbf{0.85}} & \textbf{0.61} & \multicolumn{1}{c|}{0.82} & 0.59 \\ \cline{2-13} 
\multicolumn{1}{|c|}{}                                                                                      & \multirow{5}{*}{\begin{tabular}[c]{@{}c@{}}Deep\\ Learning\\ Methods\end{tabular}}    & CNN                 & \multicolumn{1}{c|}{0.70} & 0.45 & \multicolumn{1}{c|}{0.76} & 0.46 & \multicolumn{1}{c|}{0.77} & 0.55 & \multicolumn{1}{c|}{0.80} & 0.56 & \multicolumn{1}{c|}{0.82} & 0.55 \\ \cline{3-13} 
\multicolumn{1}{|c|}{}                                                                                      &                                                                                                     & RNN                 & \multicolumn{1}{c|}{0.75} & 0.51 & \multicolumn{1}{c|}{0.80} & 0.58 & \multicolumn{1}{c|}{0.86} & 0.60 & \multicolumn{1}{c|}{0.87} & 0.62 & \multicolumn{1}{c|}{0.85} & 0.61 \\ \cline{3-13} 
\multicolumn{1}{|c|}{}                                                                                      &                                                                                                     & LSTM                & \multicolumn{1}{c|}{0.76} & 0.53 & \multicolumn{1}{c|}{0.84} & 0.59 & \multicolumn{1}{c|}{\textit{\textbf{0.93}}} & 0.63 & \multicolumn{1}{c|}{\textit{\textbf{0.93}}} & {\textit{\textbf{0.65}}} & \multicolumn{1}{c|}{0.84} & 0.59 \\ \cline{3-13} 
\multicolumn{1}{|c|}{}                                                                                      &                                                                                                     & GRU                 & \multicolumn{1}{c|}{0.75} & 0.51 & \multicolumn{1}{c|}{0.76} & 0.56 & \multicolumn{1}{c|}{0.91} & 0.62 & \multicolumn{1}{c|}{0.90} & 0.63 & \multicolumn{1}{c|}{0.86} & 0.57 \\ \cline{3-13} 
\multicolumn{1}{|c|}{}                                                                                      &                                                                                                     & TCN                 & \multicolumn{1}{c|}{0.71} & 0.49 & \multicolumn{1}{c|}{0.74} & 0.57 & \multicolumn{1}{c|}{0.88} & 0.60 & \multicolumn{1}{c|}{0.89} & 0.61 & \multicolumn{1}{c|}{0.91} & 0.59 \\ \hline
\end{tabular}

    \label{tab:table:case_1_3}
  \end{subtable}

  \label{table:case_1}
\end{table}

\newpage
\section{Distribution of time to ESRD for the ESRD cohort}
\label{app:esrd_dis}
\begin{figure}[htbp]
\begin{center}
\includegraphics[width=1\textwidth]{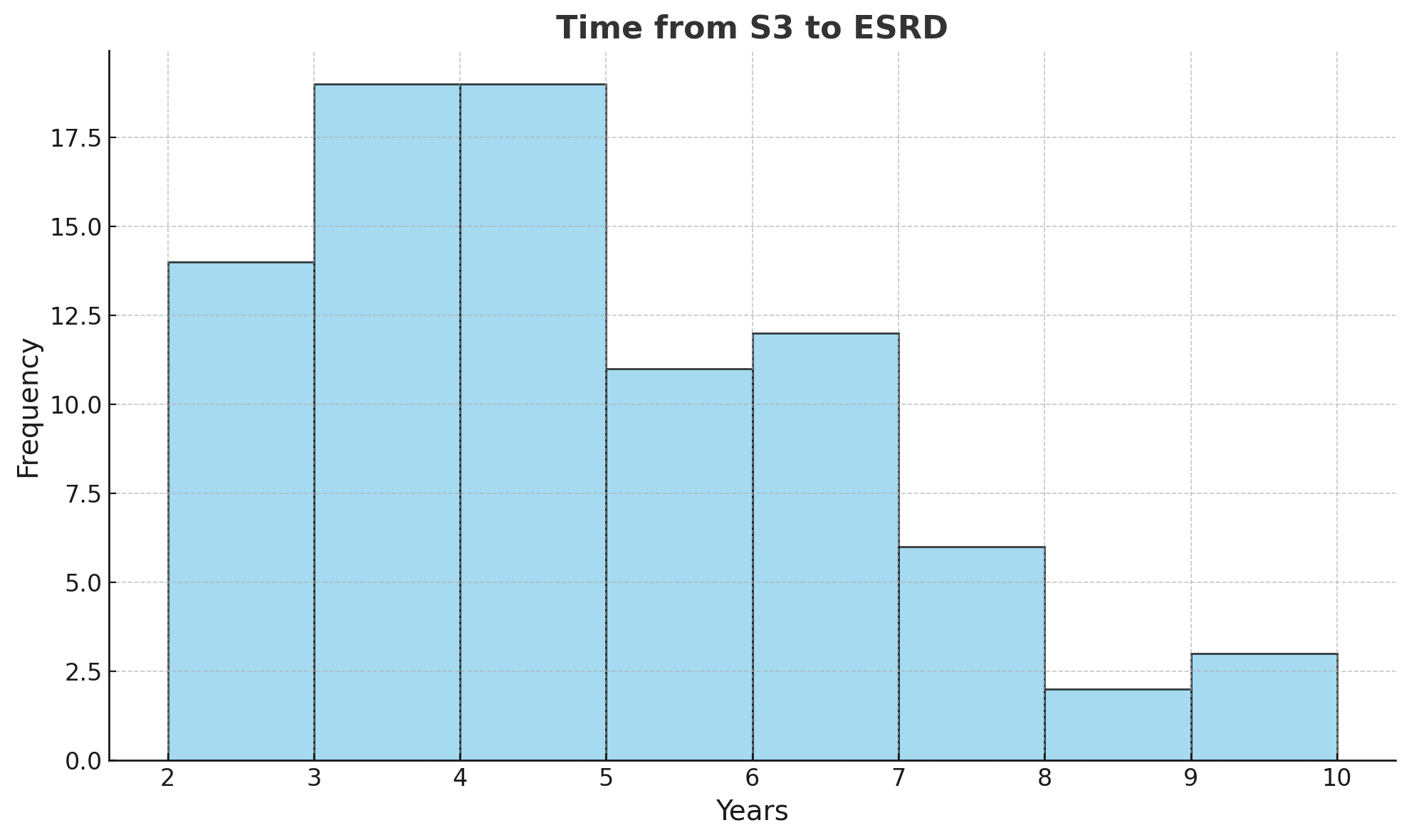}
\caption{The plot illustrates the variability in time to ESRD across the cohort, with a mean time of 4.82 years and a standard deviation of 1.82 years.}
\label{fig:S3_to_ESRD}
\end{center}
\end{figure}

\newpage
\section{Feature Abbreviations}
\label{app:features_abb}
\begin{table}[h!]
\caption{Mapping between feature abbreviations in The feature importance plot and corresponding full names.}
\centering
\label{tab:feature_names}
\begin{tabular}{|l|l|}
\hline
\textbf{Short Name} & \textbf{Full Name} \\ \hline
S5 & CKD Stage 5 \\ \hline
MA & Metabolic  Acidosis \\ \hline
Diabetes & Diabetes \\ \hline
n\_claims\_O\_min & Number of Outpatient Claims (Min) \\ \hline
Gender & Gender \\ \hline
net\_exp\_O\_std & Net Expense for Outpatient Claims (Std Dev) \\ \hline
n\_claims\_I\_std & Number of Inpatient Claims (Std Dev) \\ \hline
MI & Myocardial Infarction \\ \hline
net\_exp\_I\_max & Net Expense for Inpatient Claims (Max) \\ \hline
Athsc & Atherosclerosis \\ \hline
SH & Secondary Hyperparathyroidism \\ \hline
Phos & Phosphatemia \\ \hline
net\_exp\_O\_min & Net Expense for Outpatient Claims (Min) \\ \hline
ND & Nutritional Disorders \\ \hline
net\_exp\_DR\_max & Net Expense for Drug-Related Claims (Max) \\ \hline
Hemoglobin\_max & Hemoglobin Level (Max) \\ \hline
Serum\_Calcium\_max & Serum Calcium Level (Max) \\ \hline
Phosphorus\_std & Phosphorus Level (Std Dev) \\ \hline
Anemia & Anemia Diagnosis \\ \hline
Cvd & Cardiovascular Disease \\ \hline
Intact\_PTH\_std & Intact Parathyroid Hormone (Std Dev) \\ \hline
Prot & Proteinuria \\ \hline
Age & Age \\ \hline
Phosphorus\_max & Phosphorus Level (Max) \\ \hline
n\_claims\_DR\_min & Number of Drug-Related Claims (Min) \\ \hline
net\_exp\_P\_min & Net Expense for Pharmacy Claims (Min) \\ \hline
n\_claims\_P\_std & Number of Pharmacy Claims (Std Dev) \\ \hline
CD & Conduction \& Dysrhythmias \\ \hline
Serum\_Calcium\_min & Serum Calcium Level (Min) \\ \hline
net\_exp\_I\_std & Net Expense for Inpatient Claims (Std Dev) \\ \hline
\end{tabular}
\end{table}

\end{document}